\documentclass[11pt]{article}
\usepackage{amssymb,amsmath,amsfonts}
\usepackage{graphicx}
\usepackage{graphics}
\usepackage{eepic,epsfig}

1.60cm \makeatletter \@addtoreset{equation}{section}

\makeatother

\begin{document}

\title{Finite temperature fermionic condensate and currents \\
in topologically nontrivial spaces}
\author{S. Bellucci$^{1}$\thanks{%
E-mail: bellucci@lnf.infn.it },\, E. R. Bezerra de Mello$^{2}$\thanks{%
E-mail: emello@fisica.ufpb.br},\, A. A. Saharian$^{3}$ \thanks{%
E-mail: saharian@ysu.am} \\
\\
\textit{$^1$ INFN, Laboratori Nazionali di Frascati,}\\
\textit{Via Enrico Fermi 40, 00044 Frascati, Italy} \vspace{0.3cm}\\
\textit{$^{2}$Departamento de F\'{\i}sica, Universidade Federal da Para\'{\i}%
ba}\\
\textit{58.059-970, Caixa Postal 5.008, Jo\~{a}o Pessoa, PB, Brazil}\vspace{%
0.3cm}\\
\textit{$^3$Department of Physics, Yerevan State University,}\\
\textit{1 Alex Manoogian Street, 0025 Yerevan, Armenia}}
\maketitle

\begin{abstract}
We investigate the finite temperature fermionic condensate and the
expectation values of the charge and current densities for a massive fermion
field in a spacetime background with an arbitrary number of toroidally
compactified spatial dimensions in the presence of a non-vanishing chemical
potential. Periodicity conditions along compact dimensions are taken with
arbitrary phases and the presence of a constant gauge field is assumed. The
latter gives rise to Aharonov-Bohm-like effects on the expectation values.
They are periodic functions of magnetic fluxes enclosed by compact
dimensions with the period equal to the flux quantum. The current density
has nonzero components along compact dimensions only. Both low- and
high-temperature asymptotics of the expectation values are studied. In
particular, it has been shown that at high temperatures the current density
is exponentially suppressed. This behavior is in sharp contrast with the
corresponding asymptotic in the case of a scalar field, where the current
density linearly grows with the temperature. The features for the models in
odd dimensional spacetimes are discussed. Applications are given to
cylindrical and toroidal nanotubes described within the framework of
effective Dirac theory for the electronic subsystem.
\end{abstract}

\bigskip

PACS numbers: 03.70.+k, 11.10.Kk, 03.75.Hh, 61.46.Fg

\bigskip

\section{Introduction}

There exists a variety of different models in which the physical problem is
formulated in spacetime backgrounds having compact spatial dimensions. In
high energy physics, the well-known examples are Kaluza-Klein type models,
supergravity and superstring theories. The models with a compact universe
may play an important role in providing proper initial conditions for
inflation \cite{Lind04} (for physical motivations of considering compact
universes see also Ref. \cite{Star98}). An interesting application of the
field theoretical models having compact dimensions recently appeared in
condensed matter physics. The long-wavelength description of the electronic
states in graphene can be formulated in terms of an effective Dirac theory
in a two-dimensional space with the Fermi velocity playing the role of the
speed of light (for a review see Ref. \cite{Gusy07}). Carbon nanotubes are
generated by rolling up a graphene sheet to form a cylinder, and the
background space for the corresponding Dirac-like theory has the topology $%
R^{1}\times S^{1}$. The compactification along the nanotube axis gives rise
to another class of graphene-made structures called toroidal carbon
nanotubes having the topology of a two-torus.

In field-theoretical models formulated on a spacetime background with
compact dimensions, the periodicity conditions imposed on fields separate
configurations with suitable wavelengths. This leads to the shift in the
expectation values of various physical observables in quantum field theory.
In particular, many authors have investigated the vacuum energy and stresses
induced by the presence of compact dimensions (for reviews see Refs. \cite%
{Most97}, \cite{Duff86}). This effect, known as topological Casimir effect,
is a physical example of the connection between global properties of
spacetime and quantum phenomena. In higher-dimensional field theories with
compact extra dimensions, the Casimir energy induces an effective potential
providing a stabilization mechanism for moduli fields and thereby fixing the
effective gauge couplings. The Casimir effect has also been considered as an
origin for the dark energy in Kaluza-Klein type models and in braneworlds
\cite{DarkEn}.

The effects of the toroidal compactification of spatial dimensions on the
properties of quantum vacuum for various spin fields have been discussed by
several authors (see, for instance, Refs. \cite{Most97}-\cite{Bell09}, and
references therein). One-loop quantum effects in de Sitter spacetime with
toroidally compact dimensions are studied in Refs. \cite{Saha08} and \cite%
{Saha08b} for scalar and fermionic fields, respectively. The main part of
the papers, devoted to the influence of the nontrivial topology on the
properties of the quantum vacuum, considers the vacuum energy and stresses.
These quantities are chosen because of their close connection with the
structure of spacetime through the general theory of gravity. Another
important characteristic for charged fields is the expectation value of the
current density. In Ref.~\cite{Bell10}, we have investigated the vacuum
expectation value of the current density for a fermionic field in spaces
with an arbitrary number of toroidally compactified dimensions. Application
of the general results were given to the electrons of a graphene sheet
rolled into cylindrical and toroidal shapes. Combined effects of compact
spatial dimensions and boundaries on the vacuum expectation values of the
fermionic current have been discussed recently in Ref. \cite{Bell13}. The
geometry of boundaries is given by two parallel plates on which the fermion
field obeys bag boundary conditions. Vacuum expectation values of the
current densities for charged scalar and Dirac spinor fields in de Sitter
spacetime with toroidally compact spatial dimensions are investigated in
Ref. \cite{Bell13b}. The effects of nontrivial topology around a conical
defect on the current induced by a magnetic flux were discussed in Ref. \cite%
{Srir01} for scalar and fermion fields.

The finite temperature effects are of key importance in both types of models
with compact dimensions used in the cosmology of the early Universe and in
condensed matter physics. In Ref. \cite{Beze13}, we have investigated the
finite temperature expectation values of the charge and current densities
for a complex scalar field with nonzero chemical potential in the background
of a flat spacetime with spatial topology $R^{p}\times (S^{1})^{q}$. In the
latter, the separate contributions to the charge and current densities
coming from the Bose-Einstein condensate and from excited states were
studied. Continuing in this line of investigations, in the present paper we
consider the effects of toroidal compactification of spatial dimensions on
the finite temperature fermionic condensate, charge and current densities
for a massive field in the presence of a non-vanishing chemical potential.
The thermal Casimir effect in cosmological models with nontrivial topology
has been discussed in Refs. \cite{Alta78}. A general discussion of the
finite temperature effects for a scalar field in higher dimensional product
manifolds with compact subspaces is given in Ref. \cite{Dowk84}. Specific
calculations are presented for the cases with the internal space being a
torus or a sphere. In Ref. \cite{Campo91}, the corresponding results are
extended to the case with a nonzero chemical potential. In the previous
discussions about the effects from nontrivial topology and finite
temperature, the authors mainly consider periodicity and antiperiodicity
conditions imposed on the fields along compact dimensions. The latter
correspond to untwisted and twisted configurations of fields respectively.
In this case the current density corresponding to a conserved charge
associated with an internal symmetry vanishes. As it will be seen below, the
presence of a constant gauge field, interacting with a charged quantum
field, will induce a nontrivial phase in the periodicity conditions along
compact dimensions. As a consequence of this, nonzero components of the
current density appear along compact dimensions. This is a sort of
Aharonov-Bohm-like effect related to the nontrivial topology of the
background space.

In what follows we consider the fermionic condensate and the expectation
values of the charge and current densities in $(D+1)$-dimensional spacetime
with spatial topology $R^{p}\times (S^{1})^{q}$ (with $p+q=D$). The
corresponding results can be used in three types of models. For the first
one we have $p=3$, $q>1$, and it corresponds to the universe with
Kaluza-Klein-type extra dimensions. In these models, the currents along
compact dimensions are sources of cosmological magnetic fields. In the
second class of models $D=3$, and the results given below describe how the
properties of the universe are changed by one-loop quantum effects induced
by the compactness of spatial dimensions. Another possible range for the
applications of the results obtained in the present paper could be
graphene-made structures like cylindrical and toroidal carbon nanotubes
described within the framework of Dirac-like theory.

The paper is organized as follows. In the next section we describe the
geometry of the problem under consideration and present a complete set of
mode functions needed in the evaluation of expectation values. Then, the
fermionic condensate is investigated for a complex fermionic field in
thermal equilibrium. In Sections \ref{sec:Charge} and \ref{sec:Current}, the
expectation values of the charge and current densities are considered.
Several representations are provided and the asymptotic behaviors are
investigated in various limiting cases, including low- and high-temperature
limits. The features of the model in odd spacetime dimensions are discussed
in Section \ref{sec:OddDim}. Applications of general results to a $(2+1)$%
-dimensional model describing the long-wavelength excitations of the
electronic subsystem in a graphene sheet are given in Section \ref{sec:Nano}%
. The main results are summarized in Section \ref{sec:Conc}. Throughout the
paper, except in Section \ref{sec:Nano}, we use the units $\hbar =c=1$.

\section{Fermionic condensate}

\label{sec:FC}

Let us consider a fermionic field $\psi $ on a background of $(D+1)$%
-dimensional flat spacetime having the spatial topology $R^{p}\times
(S^{1})^{q}$, $p+q=D$. We shall denote by $\mathbf{z}=(\mathbf{z}_{p},%
\mathbf{z}_{q})$ the Cartesian coordinates, where $\mathbf{z}%
_{p}=(z^{1},\ldots ,z^{p})$ and $\mathbf{z}_{q}=(z^{p+1},\ldots ,z^{D})$
correspond to uncompactified and compactified dimensions, respectively. One
has $-\infty <z^{l}<\infty $ for $l=1,\ldots ,p$, and $0\leqslant
z^{l}\leqslant L_{l}$ for $l=p+1,\ldots ,D$, with $L_{l}$ being the length
of the $l$th compact dimension. In the presence of an external gauge field $%
A_{\mu }$, the evolution of the field is described by the Dirac equation
\begin{equation}
i\gamma ^{\mu }D_{\mu }\psi -m\psi =0\ ,\;D_{\mu }=\partial _{\mu }+ieA_{\mu
}.  \label{DirEq}
\end{equation}%
For the irreducible representation of the Clifford algebra the Dirac
matrices $\gamma ^{\mu }$ are $N\times N$ matrices with $N=2^{[(D+1)/2]}$
(the square brackets mean the integer part of the enclosed expression). We
take these matrices in the Dirac representation:
\begin{equation}
\gamma ^{0}=\left(
\begin{array}{cc}
1 & 0 \\
0 & -1%
\end{array}%
\right) ,\;\gamma ^{\mu }=\left(
\begin{array}{cc}
0 & \sigma _{\mu } \\
-\sigma _{\mu }^{+} & 0%
\end{array}%
\right) ,\;\mu =1,2,\ldots ,D.  \label{gammatrix}
\end{equation}%
From the anticommutation relations for the Dirac matrices we get $\sigma
_{\mu }\sigma _{\nu }^{+}+\sigma _{\nu }\sigma _{\mu }^{+}=2\delta _{\mu \nu
}$. In the case $D=2$ one has $N=2$ and the Dirac matrices can be taken in
the form $\gamma ^{\mu }=(\sigma _{\mathrm{P}3},i\sigma _{\mathrm{P}%
2},-i\sigma _{\mathrm{P}1})$, with $\sigma _{\mathrm{P}\mu }$ being the
usual $2\times 2$ Pauli matrices \footnote{%
Some special features of the model in odd dimensional spacetimes will be
discussed in Section \ref{sec:OddDim}}. In what follows we assume that $%
A_{\mu }=(A_{0},-\mathbf{A})$ is a constant vector potential. Although the
corresponding magnetic field strength vanishes, the non-trivial topology of
the background spacetime induces Aharonov-Bohm-like effects for expectation
values of physical observables. As it will be seen below, the fermionic
condensate (FC) and the current density are periodic functions of the
components of the gauge field along compact dimensions.

One of the characteristic features of field theory on backgrounds with
nontrivial topology is the appearance of topologically inequivalent field
configurations and, in addition to the field equation (\ref{DirEq}), we need
to specify the periodicity conditions obeyed by the field operator along
compact dimensions. Here we consider the generic quasiperiodicity boundary
conditions,%
\begin{equation}
\psi (t,\mathbf{z}_{p},\mathbf{z}_{q}+L_{l}\mathbf{e}_{l})=e^{2\pi i\alpha
_{l}}\psi (t,\mathbf{z}_{p},\mathbf{z}_{q}),  \label{PerCond}
\end{equation}%
with constant phases $\alpha _{l}$ and with $\mathbf{e}_{l}$\ being the unit
vector along the direction of the coordinate $z^{l}$, $l=p+1,\ldots ,D$.
Twisted and untwisted periodicity conditions, most often discussed in the
literature, correspond to special cases $\alpha _{l}=1/2$ and $\alpha _{l}=0$%
, respectively. As it will be discussed below, for a Dirac field describing
the long-wavelength properties of graphene, the cases $\alpha _{l}=0,\pm 1/3$
are realized in carbon nanotubes.

Here, we are interested in the effects of non-trivial topology and the
magnetic fluxes enclosed by compact dimensions on the FC and the expectation
values of the charge and current densities assuming that the field is in
thermal equilibrium at finite temperature $T$. Firstly, we consider the FC
defined as%
\begin{equation}
\left\langle \bar{\psi}\psi \right\rangle =\mathrm{tr}[\widehat{\rho }\bar{%
\psi}\psi ],  \label{FC}
\end{equation}%
where $\bar{\psi}=\psi ^{+}\gamma ^{0}$ is the Dirac conjugated spinor, $%
\hat{\rho}$ is the density matrix and $\left\langle \cdots \right\rangle $
means the ensemble average. For the thermodynamical equilibrium distribution
at temperature $T$, the density matrix has the standard form:
\begin{equation}
\hat{\rho}=Z^{-1}e^{-\beta (\hat{H}-\mu ^{\prime }\hat{Q})},\;\beta =1/T,
\label{rho}
\end{equation}%
where $\hat{H}$ is the Hamilton operator, $\widehat{Q}$ denotes a conserved
charge and $\mu ^{\prime }$ is the related chemical potential. The
grand-canonical partition function $Z$ is defined as:%
\begin{equation}
Z=\mathrm{tr}[e^{-\beta (\hat{H}-\mu ^{\prime }\hat{Q})}].  \label{PartFunc}
\end{equation}

For the evaluation of the expectation value in Eq. (\ref{FC}) we shall
employ the direct mode summation technique. Let $\{\psi _{\sigma
}^{(+)},\psi _{\sigma }^{(-)}\}$ be a complete set of normalized positive-
and negative-energy solutions of Eq. (\ref{DirEq}) obeying the
quasiperiodicity conditions (\ref{PerCond}). The corresponding energies will
be denoted by $\varepsilon _{\sigma }^{(\pm )}$. Here $\sigma $ stands for a
set of quantum numbers specifying the solutions. For the evaluation of the
FC we expand the field operator as:
\begin{equation}
\psi =\sum_{\sigma }[\hat{a}_{\sigma }\psi _{\sigma }^{(+)}+\hat{b}_{\sigma
}^{+}\psi _{\sigma }^{(-)}],  \label{psiexp}
\end{equation}%
and use the relations%
\begin{eqnarray}
\mathrm{tr}[\hat{\rho }\hat{a}_{\sigma }^{+}\hat{a}_{\sigma ^{\prime }}] &=&%
\frac{\delta _{\sigma \sigma ^{\prime }}}{e^{\beta (\varepsilon _{\sigma
}^{(+)}-\tilde{\mu })}+1},  \notag \\
\mathrm{tr}[\hat{\rho }\hat{b}_{\sigma }^{+}\hat{b}_{\sigma ^{\prime }}] &=&%
\frac{\delta _{\sigma \sigma ^{\prime }}}{e^{\beta (\varepsilon _{\sigma
}^{(-)}+\tilde{\mu })}+1} ,  \label{traa}
\end{eqnarray}%
with $\tilde{\mu }=e\mu ^{\prime }$. In Eq. (\ref{traa}), $\delta _{\sigma
\sigma ^{\prime }}$ corresponds to the Kronecker delta for the discrete
components of the collective index $\sigma $ and to the Dirac delta function
for the continuous ones. The expressions for $\mathrm{tr}[\hat{\rho }\hat{a}%
_{\sigma ^{\prime }}\hat{a}_{\sigma }^{+}]$ and $\mathrm{tr}[\hat{\rho }\hat{%
b}_{\sigma ^{\prime }}\hat{b}_{\sigma }^{+}]$ are obtained from (\ref{traa})
by using the anticommutation relations for the creation and annihilation
operators.

Substituting the expansion (\ref{psiexp}) into Eq. (\ref{FC}) and using the
relations (\ref{traa}), for the FC we get the following expression:%
\begin{equation}
\left\langle \bar{\psi}\psi \right\rangle =\left\langle \bar{\psi}\psi
\right\rangle _{0}+\sum_{\sigma }\sum_{j=+,-}\frac{j\bar{\psi}_{\sigma
}^{(j)}\psi _{\sigma }^{(j)}}{e^{\beta (\varepsilon _{\sigma }^{(j)}-j\tilde{%
\mu})}+1},  \label{FCexp}
\end{equation}%
where
\begin{equation}
\left\langle \bar{\psi}\psi \right\rangle _{0}=\sum_{\sigma }\bar{\psi}%
_{\sigma }^{(-)}\psi _{\sigma }^{(-)},  \label{FC0}
\end{equation}%
is the vacuum expectation value of the FC. The latter was investigated in
Ref. \cite{Bell09} and here we will be mainly concerned with the finite
temperature effects provided by the second term in the right-hand side of
Eq. (\ref{FCexp}). For the evaluation of the FC we need to specify the
mode-functions. In accordance with the symmetry of the problem, we take
these functions in the form of plane waves:%
\begin{eqnarray}
\psi _{\sigma }^{(+)} &=&A_{\sigma }^{(+)}e^{i\mathbf{k}\cdot \mathbf{z}%
-i\varepsilon _{\mathbf{k}}^{(+)}t}\left(
\begin{array}{c}
w_{\chi }^{(+)} \\
\frac{(\mathbf{k}-e\mathbf{A})\cdot \boldsymbol{\sigma}^{+}}{\varepsilon _{%
\mathbf{k}}^{(+)}-eA_{0}+m}w_{\chi }^{(+)}%
\end{array}%
\right) ,  \notag \\
\psi _{\sigma }^{(-)} &=&A_{\sigma }^{(-)}e^{i\mathbf{k}\cdot \mathbf{z}%
+i\varepsilon _{\mathbf{k}}^{(-)}t}\left(
\begin{array}{c}
-\frac{(\mathbf{k}-e\mathbf{A})\cdot \boldsymbol{\sigma}}{\varepsilon _{%
\mathbf{k}}^{(-)}+eA_{0}+m}w_{\chi }^{(-)} \\
w_{\chi }^{(-)}%
\end{array}%
\right) ,  \label{psiplm}
\end{eqnarray}%
where $\boldsymbol{\sigma}=(\sigma _{1},\sigma _{2},\ldots ,\sigma _{D})$
and
\begin{equation}
\varepsilon _{\mathbf{k}}^{(\pm )}=\varepsilon _{\sigma }^{(\pm )}=\pm
eA_{0}+\sqrt{\left( \mathbf{k}-e\mathbf{A}\right) ^{2}+m^{2}}.  \label{epspm}
\end{equation}%
In Eq. (\ref{psiplm}), $w_{\chi }^{(\pm )}$, $\chi =1,\ldots ,N/2$, are
one-column matrices having $N/2$ rows with the elements $w_{\chi l}^{(\pm
)}=\delta _{\chi l}$.

For the components of the momentum $\mathbf{k}$ along uncompactified
dimensions one has $-\infty <k_{l}<\infty $, $l=1,\ldots ,p$. The components
along compact dimensions are quantized by the quasiperiodicity conditions (%
\ref{PerCond}) and the corresponding eigenvalues are given by%
\begin{equation}
k_{l}=2\pi (n_{l}+\alpha _{l})/L_{l},\;n_{l}=0,\pm 1,\pm 2,\ldots ,
\label{klcomp}
\end{equation}%
with $l=p+1,\ldots ,D$. The coefficients $A_{\sigma }^{(\pm )}$ in Eq. (\ref%
{psiplm}) are determined by the orthonormalization condition $\int
d^{D}x\,\psi _{\sigma }^{(\lambda )+}\psi _{\sigma ^{\prime }}^{(\lambda
^{\prime })}=\delta _{\sigma \sigma ^{\prime }}\delta _{\lambda \lambda
^{\prime }}$ with $\lambda ,\lambda ^{\prime }=+,-$. This gives:%
\begin{equation}
A_{\sigma }^{(\pm )2}=\frac{\varepsilon _{\mathbf{k}}^{(\pm )}\mp eA_{0}+m}{%
2(2\pi )^{p}V_{q}(\varepsilon _{\mathbf{k}}^{(\pm )}\mp eA_{0})},
\label{Asig}
\end{equation}%
where $V_{q}=L_{p+1}\cdots L_{D}$ is the volume of the compact subspace. Now
the set of quantum numbers is specified to $\sigma =(\mathbf{k}_{p},\mathbf{n%
}_{q},\chi )$, where $\mathbf{k}_{p}=(k_{1},\ldots ,k_{p})$, $\mathbf{n}%
_{q}=(n_{p+1},\ldots ,n_{D})$, and%
\begin{equation}
\sum_{\sigma }=\int d\mathbf{k}_{p}\sum_{\mathbf{n}_{q}\in \mathbf{Z}%
^{q}}\sum_{\chi =1}^{N/2}.
\end{equation}

Substituting the mode functions (\ref{psiplm}) into Eq. (\ref{FCexp}) and
shifting the integration variable $\mathbf{k}_{p}\rightarrow \mathbf{k}_{p}+e%
\mathbf{A}_{p}$, we find the following expression:%
\begin{equation}
\left\langle \bar{\psi}\psi \right\rangle =\left\langle \bar{\psi}\psi
\right\rangle _{0}+\sum_{j=+,-}\left\langle \bar{\psi}\psi \right\rangle
_{j},  \label{FCdec}
\end{equation}%
where%
\begin{equation}
\left\langle \bar{\psi}\psi \right\rangle _{\pm }=\frac{Nm}{2(2\pi )^{p}V_{q}%
}\int d\mathbf{k}_{p}\sum_{\mathbf{n}_{q}\in \mathbf{Z}^{q}}\frac{%
1/\varepsilon (\mathbf{k})}{e^{\beta (\varepsilon (\mathbf{k})\mp \mu )}+1},
\label{FCpm}
\end{equation}%
is the part in the FC coming from particles (upper sign) and antiparticles
(lower sign). In Eq. (\ref{FCpm}), $\mu =\tilde{\mu }-eA_{0}$,%
\begin{equation}
\varepsilon (\mathbf{k})=\sqrt{\mathbf{k}_{p}^{2}+\varepsilon _{\mathbf{n}%
_{q}}^{2}},\;\varepsilon _{\mathbf{n}_{q}}=\sqrt{\sum\nolimits_{l=p+1}^{D}%
\tilde{k}_{l}^{2}+m^{2}},  \label{epsk}
\end{equation}%
and we have introduced the notations%
\begin{equation}
\tilde{k}_{l}=2\pi (n_{l}+\tilde{\alpha}_{l})/L_{l},\;\tilde{\alpha}%
_{l}=\alpha _{l}-eL_{l}\mathbf{A}_{l}/2\pi .  \label{epsn}
\end{equation}%
As it is seen, the FC does not depend on the components of the vector
potential along non compact dimensions.

The dependence on the phases $\alpha _{l}$ and on the components of the
vector potential along compact dimensions enters in the form of the
combination $\tilde{\alpha}_{l}$ in Eq. (\ref{epsn}). We could see this
directly by the gauge \ transformation $A_{\mu }=A_{\mu }^{\prime }+\partial
_{\mu }\Lambda (x)$, $\psi (x)=\psi ^{\prime }(x)e^{-ie\Lambda (x)}$, with $%
\Lambda (x)=A_{\mu }z^{\mu }$. The new function $\psi ^{\prime }(x)$ obeys
the Dirac equation with $A_{\mu }^{\prime }=0$ and the periodicity condition
$\psi ^{\prime }(t,\mathbf{z}_{p},\mathbf{z}_{q}+L_{l}\mathbf{e}%
_{l})=e^{2\pi i\tilde{\alpha}_{l}}\psi ^{\prime }(t,\mathbf{z}_{p},\mathbf{z}%
_{q})$. The expectation values are not changed under the gauge
transformation and, in the new gauge, the parameter $\tilde{\alpha}_{l}$
appears, instead of $\alpha _{l}$ and $\mathbf{A}_{l}$. Note that we can
write $eL_{l}\mathbf{A}_{l}/2\pi =\phi _{l}/\phi _{0}$, where $\phi _{l}$ is
the magnetic flux enclosed by the $l$th compact dimensions and $\phi
_{0}=2\pi /e$ is the flux quantum. Hence, the presence of a constant gauge
field is equivalent to the shift of the phases in the periodicity conditions
along compact dimensions. In particular, a nontrivial phase is induced for
the special cases of twisted and untwisted fermionic fields. As it will be
discussed below, this leads to the appearance of the nonzero current density
along compact dimensions. Another interesting physical effect of a constant
gauge field is the topological generation of a gauge field mass by the
toroidal spacetime (see Ref. \cite{Acto90} and references therein).

If we present the phases $\tilde{\alpha}_{l}$ in the form $\tilde{\alpha}%
_{l}=\tilde{\alpha}_{l}^{(f)}+\tilde{\alpha}_{l}^{(i)}$, with $\tilde{\alpha}%
_{l}^{(i)}$ being an integer and $|\tilde{\alpha}_{l}^{(f)}|\leqslant 1/2$,
then, as it is seen from Eq. (\ref{FCpm}), the FC depends on the fractional
part $\tilde{\alpha}_{l}^{(f)}$ only. We will denote by $\varepsilon _{0}$
the smallest value of the energy $\varepsilon (\mathbf{k})$:%
\begin{equation}
\varepsilon _{0}=\sqrt{\sum\nolimits_{l=p+1}^{D}(2\pi \tilde{\alpha}%
_{l}^{(f)}/L_{l})^{2}+m^{2}},  \label{epsmin}
\end{equation}%
which corresponds to $n_{l}=-\tilde{\alpha}_{l}^{(i)}$. Note that, while the
bosonic chemical potential is restricted by $|\mu |\leqslant $ $\varepsilon
_{0}$, the fermionic chemical potential can have any value with respect to $%
\varepsilon _{0}$.

Firstly, we consider the case $|\mu |<$ $\varepsilon _{0}$. With this
assumption, by making use of the expansion%
\begin{equation}
(e^{y}+1)^{-1}=-\sum_{n=1}^{\infty }(-1)^{n}e^{-ny} \ ,  \label{Expansion}
\end{equation}%
after the integration over $\mathbf{k}_{p}$, from (\ref{FCpm}) we get%
\begin{equation}
\left\langle \bar{\psi}\psi \right\rangle _{\pm }=-\frac{Nm}{(2\pi
)^{(p+1)/2}V_{q}}\sum_{n=1}^{\infty }(-1)^{n}e^{\pm n\beta \mu }\sum_{%
\mathbf{n}_{q}\in \mathbf{Z}^{q}}\varepsilon _{\mathbf{n}%
_{q}}^{p-1}f_{(p-1)/2}(n\beta \varepsilon _{\mathbf{n}_{q}}).  \label{FCpm1}
\end{equation}%
In (\ref{FCpm1}) and in what follows we use the notation%
\begin{equation}
f_{\nu }(x)=K_{\nu }(x)/x^{\nu },  \label{fnju}
\end{equation}%
with $K_{\nu }(x)$ being the Macdonald function. Combining Eqs. (\ref{FCdec}%
) and (\ref{FCpm1}), for the total FC one finds:%
\begin{equation}
\left\langle \bar{\psi}\psi \right\rangle =\left\langle \bar{\psi}\psi
\right\rangle _{0}-\frac{2Nm}{(2\pi )^{(p+1)/2}V_{q}}\sum_{n=1}^{\infty
}(-1)^{n}\cosh (n\beta \mu )\sum_{\mathbf{n}_{q}\in \mathbf{Z}%
^{q}}\varepsilon _{\mathbf{n}_{q}}^{p-1}f_{(p-1)/2}(n\beta \varepsilon _{%
\mathbf{n}_{q}}).  \label{FCInt1}
\end{equation}%
As it is seen from the above equation, the FC is a periodic function of $%
\tilde{\alpha}_{l}$ with the period equal to unity. In particular, FC\ is a
periodic function of the flux enclosed by a compact dimension with the
period equal to the flux quantum.

Let us consider the behavior of the FC, given by Eq. (\ref{FCInt1}), in the
asymptotic regions of the parameters. If the lengths of the compact
dimensions are large, $L_{l}\gg \beta ,m$, the dominant contribution comes
from large values of $|n_{l}|$, and we can replace the summation by the
integration, $\sum_{\mathbf{n}_{q}\in \mathbf{Z}^{q}}\rightarrow V_{q}(2\pi
)^{-q}\int d\mathbf{k}_{q}$ with $\varepsilon _{\mathbf{n}_{q}}=\sqrt{%
\mathbf{k}_{q}^{2}+m^{2}}$. The integral is evaluated by using the formula%
\begin{equation}
\int d^{r}\mathbf{x}\,(\mathbf{x}^{2}+a^{2})^{\nu }f_{\nu }(b\sqrt{\mathbf{x}%
^{2}+a^{2}})=(2\pi )^{r/2}a^{2\nu +r}f_{\nu +r/2}(ab),  \label{IntFormu}
\end{equation}%
and, to the leading order, we obtain the FC in topologically trivial
Minkowski spacetime:%
\begin{equation}
\left\langle \bar{\psi}\psi \right\rangle _{\mathrm{M}}=-\frac{2Nm^{D}}{%
(2\pi )^{(D+1)/2}}\sum_{n=1}^{\infty }(-1)^{n}\cosh (n\beta \mu
)f_{(D-1)/2}(mn\beta ).  \label{FCM}
\end{equation}%
Here we have renormalized the zero temperature FC in Minkowski spacetime to
zero. More precisely, if for a part of compact dimensions (with the index $r$%
) the lengths are large, by replacing the corresponding series by the
integrations and using (\ref{IntFormu}), it is seen from Eq. (\ref{FCInt1})
that the result obtained is equivalent to that for the topology $%
R^{p+r}\times (S^{1})^{q-r}$.

In the opposite limit, when the length of the $l$th compact dimension is
small compared with the other length scales and $L_{l}\ll \beta $, under the
assumption $|\tilde{\alpha}_{l}|<1/2$, the dominant contribution to the
series over $n_{l}$ in Eq. (\ref{FCInt1}) comes from the term $n_{l}=0$. It
can be seen that for $\tilde{\alpha}_{l}=0$, in the leading order, $%
L_{l}\left\langle \bar{\psi}\psi \right\rangle $ coincides with $%
N\left\langle \bar{\psi}\psi \right\rangle _{(p,q-1)}/N_{D-1}$, where $%
\left\langle \bar{\psi}\psi \right\rangle _{(p,q-1)}$ is the condensate in $%
(D-1)$-dimensional space of topology $R^{p}\times (S^{1})^{q-1}$ with the
lengths of the compact dimensions $L_{p+1}$,\ldots ,$L_{l-1}$,$L_{l+1}$%
,\ldots ,$L_{D}$ ($N_{D-1}$ is the corresponding number of spinor
components). For $\tilde{\alpha}_{l}\neq 0$ and for small values of $L_{l}$,
the argument of the Macdonald function in Eq. (\ref{FCInt1}) is large and
the FC is suppressed by the factor $e^{-2\pi |\tilde{\alpha}_{l}|\beta
/L_{l}}$.

In the low-temperature limit, the parameter $\beta $ is large and the
dominant contribution to the FC comes from the term $n=1$ in the series over
$n$, and from the term in the series over $\mathbf{n}_{q}$ with the smallest
value of $\varepsilon _{\mathbf{n}_{q}}$. In the leading order we find%
\begin{equation}
\left\langle \bar{\psi}\psi \right\rangle \approx \left\langle \bar{\psi}%
\psi \right\rangle _{0}+\frac{Nm\varepsilon _{0}^{p/2-1}e^{-\beta
(\varepsilon _{0}-|\mu |)}}{2(2\pi )^{p/2}V_{q}\beta ^{p/2}},  \label{FClow}
\end{equation}%
with $\varepsilon _{0}$ given by Eq. (\ref{epsmin}).

An alternative expression for the FC can be obtained by employing the zeta
function technique \cite{Kirs01}. We write the expression (\ref{FCdec}) in
the form%
\begin{equation}
\left\langle \bar{\psi}\psi \right\rangle =-\frac{Nm}{(2\pi )^{p}V_{q}}\int d%
\mathbf{k}_{p}\sum_{\mathbf{n}_{q}\in \mathbf{Z}^{q}}\sideset{}{'}{\sum}%
_{n=0}^{\infty }(-1)^{n}\cosh (n\beta \mu )\frac{e^{-n\beta \varepsilon (%
\mathbf{k})}}{\varepsilon (\mathbf{k})},  \label{FCzeta}
\end{equation}%
where the prime on the sign of the sum means that the term $n=0$ should be
taken with the coefficient 1/2. This term corresponds to the vacuum
expectation value of the FC, $\left\langle \bar{\psi}\psi \right\rangle _{0}$%
. Next, we use the integral representation
\begin{equation}
\frac{e^{-n\beta \varepsilon (\mathbf{k})}}{\varepsilon (\mathbf{k})}=\frac{2%
}{\sqrt{\pi }}\int_{0}^{\infty }ds\,e^{-\varepsilon ^{2}(\mathbf{k}%
)s^{2}-n^{2}\beta ^{2}/4s^{2}},  \label{IntRep}
\end{equation}%
and the formula%
\begin{equation}
\sideset{}{'}{\sum}_{n=0}^{\infty }(-1)^{n}\cosh (n\beta \mu )e^{-n^{2}\beta
^{2}/4s^{2}}=s\frac{\sqrt{\pi }}{\beta }\sum_{n=-\infty }^{+\infty
}e^{-s^{2}[2\pi (n+1/2)/\beta +i\mu ]^{2}}.  \label{Sum1}
\end{equation}%
The above result can be proved by making use of the Poisson summation
formula. As a result, after the integration over $s$, the FC\ is presented as%
\begin{equation}
\left\langle \bar{\psi}\psi \right\rangle =-Nm\zeta (1),  \label{FCzet1}
\end{equation}%
where we have defined the zeta function as shown below:
\begin{equation}
\zeta (s)=\frac{1}{V_{q}\beta }\int \frac{d\mathbf{k}_{p}}{(2\pi )^{p}}\sum_{%
\mathbf{n}_{q+1}\in \mathbf{Z}^{q+1}}\Big\{\mathbf{k}_{p}^{2}+%
\sum_{l=p+1}^{D+1}[2\pi (n_{l}+\tilde{\alpha}_{l})/L_{l}]^{2}+m^{2}\Big\}%
^{-s},  \label{Zeta}
\end{equation}%
with $\mathbf{n}_{q+1}=(\mathbf{n}_{q},n_{D+1})$,
\begin{equation}
L_{D+1}=\beta ,\;\tilde{\alpha}_{D+1}=\frac{1}{2}+\frac{i\beta \mu }{2\pi }.
\label{LD1}
\end{equation}

The renormalized value for the FC is obtained by the analytical continuation
of the zeta function at $s=1$. An exponentially convergent expression for
the analytic continuation is obtained by integrating over $\mathbf{k}_{p}$
and then applying the generalized Chowla-Selberg formula \cite{Eliz98}. In
this way we get:
\begin{eqnarray}
\zeta (s) &=&\frac{m^{D+1-2s}}{(4\pi )^{(D+1)/2}}\frac{\Gamma (s-(D+1)/2)}{%
\Gamma (s)}+\frac{2^{1-s}m^{D+1-2s}}{(2\pi )^{(D+1)/2}\Gamma (s)}  \notag \\
&&\times \sideset{}{'}{\sum}_{\mathbf{m}_{q+1}\in \mathbf{Z}^{q+1}}\cos
(2\pi \mathbf{m}_{q+1}\cdot \boldsymbol{\alpha }_{q+1})f_{(D+1)/2-s}(mg(%
\mathbf{L}_{q+1},\mathbf{m}_{q+1})),  \label{Zeta1}
\end{eqnarray}%
with $\mathbf{L}_{q+1}=(L_{p+1},\ldots ,L_{D+1})$, $\boldsymbol{\alpha }%
_{q+1}=(\tilde{\alpha }_{p+1},\ldots ,\tilde{\alpha }_{D+1})$, and the prime
on the summation sign means that the term $\mathbf{m}_{q+1}=0$ should be
excluded from the sum. In Eq. (\ref{Zeta1}) we have introduced the notation%
\begin{equation}
g(\mathbf{L}_{q+1},\mathbf{m}_{q+1})=\left(
\sum\nolimits_{i=p+1}^{D+1}L_{i}^{2}m_{i}^{2}\right) ^{1/2}.  \label{g}
\end{equation}%
The contribution of the first term in the right-hand side of Eq. (\ref{Zeta1}%
) gives the zero temperature FC in the topologically trivial Minkowski
spacetime. This part is subtracted in the renormalization procedure. The
remaining part of the zeta function is finite, at the physical point $s=1$
and for the renormalized FC one finds%
\begin{equation}
\left\langle \bar{\psi}\psi \right\rangle =-\frac{Nm^{D}}{(2\pi )^{(D+1)/2}}%
\sideset{}{'}{\sum}_{\mathbf{m}_{q+1}\in \mathbf{Z}^{q+1}}\cos (2\pi \mathbf{%
m}_{q+1}\cdot \boldsymbol{\alpha }_{q+1})f_{(D-1)/2}(mg(\mathbf{L}_{q+1},%
\mathbf{m}_{q+1})).  \label{FC2}
\end{equation}%
In this formula, the term $m_{D+1}=0$ corresponds to the vacuum expectation
value of the FC:
\begin{equation}
\left\langle \bar{\psi}\psi \right\rangle _{0}=-\frac{Nm^{D}}{(2\pi
)^{(D+1)/2}}\sideset{}{'}{\sum}_{\mathbf{m}_{q}\in \mathbf{Z}^{q}}\cos (2\pi
\mathbf{m}_{q}\cdot \boldsymbol{\alpha }_{q})f_{(D-1)/2}(mg(\mathbf{L}_{q},%
\mathbf{m}_{q})),  \label{FC01}
\end{equation}%
where, again, the prime means that the term with $\mathbf{m}_{q}=0$ should
be excluded.

Extracting the vacuum expectation value, the renormalized FC can also be
written in the form:
\begin{eqnarray}
\left\langle \bar{\psi}\psi \right\rangle &=&\left\langle \bar{\psi}\psi
\right\rangle _{0}-\frac{2Nm^{D}}{(2\pi )^{(D+1)/2}}\sum_{n=1}^{\infty
}(-1)^{n}\cosh (n\beta \mu )  \notag \\
&&\times \sum_{\mathbf{m}_{q}\in \mathbf{Z}^{q}}\cos (2\pi \mathbf{m}%
_{q}\cdot \boldsymbol{\alpha }_{q})f_{(D-1)/2}(m\sqrt{g^{2}(\mathbf{L}_{q},%
\mathbf{m}_{q})+n^{2}\beta ^{2}}),  \label{FC3}
\end{eqnarray}%
with the notation%
\begin{equation}
g(\mathbf{L}_{q},\mathbf{m}_{q})=\left(
\sum\nolimits_{i=p+1}^{D}L_{i}^{2}m_{i}^{2}\right) ^{1/2}.  \label{gD}
\end{equation}%
The equivalence of two representations, given by Eqs. (\ref{FCInt1}) and (%
\ref{FC3}), is seen by using the formula (see, for instance, Ref. \cite%
{Bell09})%
\begin{equation}
\sum_{\mathbf{m}_{q}\in \mathbf{Z}^{q}}\cos (2\pi \mathbf{m}_{q}\cdot %
\boldsymbol{\alpha }_{q})f_{\nu }(m\sqrt{g^{2}(\mathbf{L}_{q},\mathbf{m}%
_{q})+n^{2}\beta ^{2}})=\frac{(2\pi )^{q/2}}{V_{q}m^{2\nu }}\sum_{\mathbf{n}%
_{q}\in \mathbf{Z}^{q}}\varepsilon _{\mathbf{n}_{q}}^{2\nu -q}f_{\nu
-q/2}(n\beta \varepsilon _{\mathbf{n}_{q}}).  \label{FormulEquiv}
\end{equation}

In order to investigate the high-temperature asymptotic of the FC, it is
convenient to transform the expression (\ref{FC2}). In this expression, the
term with $\mathbf{m}_{q}=0$ corresponds to the FC in the topologically
trivial Minkowski spacetime (see Eq. (\ref{FCM})). Separating this part from
the right-hand side of Eq. (\ref{FC2}), in the remaining part instead of $%
\cos (2\pi \mathbf{m}_{q+1}\cdot \boldsymbol{\alpha
}_{q+1})$ we write $\cos (2\pi \mathbf{m}_{q}\cdot \boldsymbol{\alpha }%
_{q})\cos (2\pi m_{D+1}\tilde{\alpha}_{D+1})$. Next, to the series over $%
m_{D+1}$ we apply the formula%
\begin{eqnarray}
&&\sum_{m_{D+1}=-\infty }^{+\infty }\cos (2\pi m_{D+1}\tilde{\alpha}%
_{D+1})f_{\nu }(m\sqrt{\beta ^{2}m_{D+1}^{2}+a^{2}})=\frac{(2\pi )^{1/2}}{%
\beta m^{2\nu }}  \notag \\
&&\quad \times \sum_{n=-\infty }^{+\infty }\left[ (2\pi (n+\tilde{\alpha}%
_{D+1})/\beta )^{2}+m^{2}\right] ^{\nu -1/2}f_{\nu -1/2}(a\sqrt{(2\pi (n+%
\tilde{\alpha}_{D+1})/\beta )^{2}+m^{2}}),  \label{SumHighT}
\end{eqnarray}%
which is a special case of Eq. (\ref{FormulEquiv}). This leads to the
representation%
\begin{eqnarray}
\left\langle \bar{\psi}\psi \right\rangle &=&\left\langle \bar{\psi}\psi
\right\rangle _{\mathrm{M}}-\frac{NmT}{(2\pi )^{D/2}}\sideset{}{'}{\sum}_{%
\mathbf{m}_{q}\in \mathbf{Z}^{q}}\cos (2\pi \mathbf{m}_{q}\cdot %
\boldsymbol{\alpha }_{q})  \notag \\
&&\times \sum_{n=-\infty }^{+\infty }\left[ (\pi \left( 2n+1\right) T+i\mu
)^{2}+m^{2}\right] ^{D/2-1}  \notag \\
&&\times f_{D/2-1}(g(\mathbf{L}_{q},\mathbf{m}_{q})\sqrt{(\pi \left(
2n+1\right) T+i\mu )^{2}+m^{2}}).  \label{FC3b}
\end{eqnarray}%
At high temperatures, the dominant contribution comes from the terms with $%
n=0$ and $n=-1$ and from the terms in the sum over $\mathbf{m}_{q}$ with the
smallest value of $g(\mathbf{L}_{q},\mathbf{m}_{q})$. If $L_{\mathrm{min}}=%
\mathrm{min}(L_{p+1},\ldots ,L_{D})$, then, to the leading order, one finds%
\begin{equation}
\left\langle \bar{\psi}\psi \right\rangle \approx \left\langle \bar{\psi}%
\psi \right\rangle _{\mathrm{M}}-\frac{2Nme^{-\pi L_{\mathrm{min}}T}}{\pi
(2\beta L_{\mathrm{min}})^{(D-1)/2}}\sum_{l}\cos (2\pi \tilde{\alpha}_{l}),
\label{FChighT}
\end{equation}%
where the summation goes over the compact dimensions with $L_{l}=L_{\mathrm{%
min}}$. From Eq. (\ref{FChighT}) we conclude that at high temperatures the
topological part in the FC is exponentially suppressed. At high
temperatures, for the Minkowskian part, to the leading order, we have%
\begin{equation}
\left\langle \bar{\psi}\psi \right\rangle _{\mathrm{M}}\approx \frac{%
1-2^{2-D}}{2\pi ^{(D+1)/2}}\Gamma ((D-1)/2)\zeta _{\mathrm{R}}(D-1)NmT^{D-1},
\label{FCMhighT}
\end{equation}%
where $\zeta _{\mathrm{R}}(x)$ is the Riemann zeta function. For $D=2$ one
gets $\left\langle \bar{\psi}\psi \right\rangle _{\mathrm{M}}\approx (2\pi
)^{-1}NmT\ln 2$.

An alternative representation for the FC, containing more detailed
information, can be found from Eq. (\ref{FCzeta}) by applying to the sum
over $n_{l}$, for $p+1\leqslant l\leqslant D$, the Abel-Plana summation
formula in the form \cite{Bell09,Beze08}
\begin{eqnarray}
&&\frac{2\pi }{L_{l}}\sum_{n_{l}=-\infty }^{\infty }g(\tilde{k}_{l})f(|%
\tilde{k}_{l}|)=\int_{0}^{\infty }dx\,[g(x)+g(-x)]f(x)+  \notag \\
&&\quad +i\int_{0}^{\infty }dx\,[f(ix)-f(-ix)]\sum_{\lambda =\pm 1}\frac{%
g(i\lambda x)}{e^{xL_{l}+2\lambda \pi i\tilde{\alpha}_{l}}-1},  \label{AP}
\end{eqnarray}%
where $\tilde{k}_{l}$ is defined by Eq. (\ref{epsn}). Taking
\begin{equation}
g(x)=1,\;f(x)=\frac{e^{-n\beta \sqrt{x^{2}+\mathbf{k}_{p}^{2}+\varepsilon _{%
\mathbf{n}_{q-1}^{l}}^{2}}}}{\sqrt{x^{2}+\mathbf{k}_{p}^{2}+\varepsilon _{%
\mathbf{n}_{q-1}^{l}}^{2}}},  \label{gf}
\end{equation}%
with $\mathbf{n}_{q-1}^{l}=(n_{p+1},\ldots ,n_{l-1},n_{l+1},\ldots ,n_{D})$,
and%
\begin{equation}
\varepsilon _{\mathbf{n}_{q-1}^{l}}=\sqrt{\varepsilon _{\mathbf{n}_{q}}^{2}-%
\tilde{k}_{l}^{2}},  \label{epsn-1}
\end{equation}%
it can be seen that the part in the FC corresponding to the first term in
the right-hand side of (\ref{AP}) corresponds to the condensate for a $D$%
-dimensional space of topology $R^{p+1}\times (S^{1})^{q-1}$ with the
lengths of the compact dimensions $L_{p+1},\ldots ,L_{l-1},L_{l+1},\ldots
,L_{D}$. The latter will be denoted by $\left\langle \bar{\psi}\psi
\right\rangle _{p+1,q-1}$. In the part of the FC corresponding to the second
term in the right-hand side of Eq. (\ref{AP}) we use the expansion%
\begin{equation}
\sum_{\lambda =\pm 1}\frac{1}{e^{xL_{l}+2\lambda \pi i\tilde{\alpha}_{l}}-1}%
=2\sum_{r=1}^{\infty }e^{-xrL_{l}}\cos (2\pi r\tilde{\alpha}_{l}).
\label{ExpAP}
\end{equation}%
After the integrations over $x$ and $\mathbf{k}_{p}$ with the help of the
formula%
\begin{equation}
\int_{a}^{\infty }du\,\,u^{2\nu +1}(u^{2}-a^{2})^{\beta -1}f_{\nu
}(cu)=2^{\beta -1}a^{2\beta +2\nu }\Gamma (\beta )f_{\nu +\beta }(ac),
\label{IntForm2}
\end{equation}%
we get (the prime means that the term $n=0$ should be halved)%
\begin{eqnarray}
\left\langle \bar{\psi}\psi \right\rangle &=&\left\langle \bar{\psi}\psi
\right\rangle _{p+1,q-1}-\frac{4NmL_{l}}{(2\pi )^{p/2+1}V_{q}}%
\sideset{}{'}{\sum}_{n=0}^{\infty }(-1)^{n}\cosh (n\beta \mu )  \notag \\
&&\times \sum_{r=1}^{\infty }\cos (2\pi r\tilde{\alpha}_{l})\sum_{\mathbf{n}%
_{q-1}^{l}\in \mathbf{Z}^{q-1}}\varepsilon _{\mathbf{n}%
_{q-1}^{l}}^{p}f_{p/2}(\varepsilon _{\mathbf{n}_{q-1}^{l}}\sqrt{%
r^{2}L_{l}^{2}+n^{2}\beta ^{2}}),  \label{FC4}
\end{eqnarray}%
where the second term in the right-hand side of Eq. (\ref{FC4}) is induced
by the compactification of the $l$th dimension. In the special case of a
single compact dimension one has $l=D$, $\varepsilon _{\mathbf{n}%
_{q-1}^{l}}=m$, and from Eq. (\ref{FC4}) we obtain:%
\begin{eqnarray}
\left\langle \bar{\psi}\psi \right\rangle &=&-\frac{4Nm^{D}}{(2\pi
)^{(D+1)/2}}\sideset{}{'}{\sum}_{n=0}^{\infty }(-1)^{n}\cosh (n\beta \mu )
\notag \\
&&\times \sum_{r=1}^{\infty }\cos (2\pi r\tilde{\alpha}_{D})f_{(D-1)/2}(m%
\sqrt{r^{2}L_{D}^{2}+n^{2}\beta ^{2}}).  \label{FC1comp}
\end{eqnarray}%
This result coincides with the corresponding one obtained from Eq. (\ref{FC3}%
).

In the discussion above we have assumed that $|\mu |<$ $\varepsilon _{0}$.
Now we consider the case $|\mu |>$ $\varepsilon _{0}$. Let us denote by $%
\varepsilon _{\mathbf{n}_{q}^{(0)}}$ the largest energy for which $|\mu |>$ $%
\varepsilon _{\mathbf{n}_{q}^{(0)}}$. From Eqs. (\ref{FCdec}) and (\ref{FCpm}%
), after the integration over the angular part of $\mathbf{k}_{p}$, we get%
\begin{equation}
\left\langle \bar{\psi}\psi \right\rangle =\left\langle \bar{\psi}\psi
\right\rangle _{0}+\frac{(4\pi )^{-p/2}Nm}{\Gamma (p/2)V_{q}}%
\sum_{j=+,-}\sum_{\mathbf{n}_{q}\in \mathbf{Z}^{q}}\int_{\varepsilon _{%
\mathbf{n}_{q}}}^{\infty }dx\,\frac{(x^{2}-\varepsilon _{\mathbf{n}%
_{q}}^{2})^{p/2-1}}{e^{\beta (x-j\mu )}+1}.  \label{FCn}
\end{equation}%
The contribution of the states with the energies $\varepsilon _{\mathbf{n}%
_{q}}>|\mu |$ to the FC is treated in a way similar to that described above.
The FC\ is an even function of the chemical potential and for definiteness
we assume that $\mu >0$. In this case the finite temperature part in the FC
coming from the antiparticles remains the same, whereas in the part coming
from the particles the spectral ranges $\varepsilon _{\mathbf{n}%
_{q}}\leqslant \varepsilon _{\mathbf{n}_{q}^{(0)}}$ and $\varepsilon _{%
\mathbf{n}_{q}}>\varepsilon _{\mathbf{n}_{q}^{(0)}}$ should be treated
separately. In this way we find the following representation:
\begin{eqnarray}
\left\langle \bar{\psi}\psi \right\rangle &=&\left\langle \bar{\psi}\psi
\right\rangle _{0}+\left\langle \bar{\psi}\psi \right\rangle _{-}+\frac{%
(4\pi )^{-p/2}Nm}{\Gamma (p/2)V_{q}}\sum_{\varepsilon _{\mathbf{n}%
_{q}}\leqslant \varepsilon _{\mathbf{n}_{q}^{(0)}}}\int_{\varepsilon _{%
\mathbf{n}_{q}}}^{\infty }dx\,\frac{(x^{2}-\varepsilon _{\mathbf{n}%
_{q}}^{2})^{p/2-1}}{e^{\beta (x-\mu )}+1}  \notag \\
&&-\frac{Nm}{(2\pi )^{(p+1)/2}V_{q}}\sum_{n=1}^{\infty }(-1)^{n}e^{n\beta
\mu }\sum_{\varepsilon _{\mathbf{n}_{q}}>\varepsilon _{\mathbf{n}%
_{q}^{(0)}}}\varepsilon _{\mathbf{n}_{q}}^{p-1}f_{(p-1)/2}(n\beta
\varepsilon _{\mathbf{n}_{q}}),  \label{FCmu}
\end{eqnarray}%
where $\left\langle \bar{\psi}\psi \right\rangle _{-}$ is given by Eq. (\ref%
{FCpm1}). In the limit $T\rightarrow 0$ we get%
\begin{equation}
\left\langle \bar{\psi}\psi \right\rangle =\left\langle \bar{\psi}\psi
\right\rangle _{0}+\frac{(4\pi )^{-p/2}Nm}{\Gamma (p/2)V_{q}}%
\sum_{\varepsilon _{\mathbf{n}_{q}}\leqslant \varepsilon _{\mathbf{n}%
_{q}^{(0)}}}\int_{\varepsilon _{\mathbf{n}_{q}}}^{\mu
}dx\,(x^{2}-\varepsilon _{\mathbf{n}_{q}}^{2})^{p/2-1}.  \label{FCmut0}
\end{equation}%
The second contribution in the right-hand side comes from the particles
which occupy the states with $\varepsilon _{\mathbf{n}_{q}}<\mu $. At high
temperatures, the leading term coincides with that in the topologically
trivial Minkowski spacetime and is given by Eq. (\ref{FCMhighT}).

\section{Charge density}

\label{sec:Charge}

In this and in the following sections we shall investigate the
expectation value of the fermionic current density, assuming that
the field is in a thermal equilibrium at finite temperature $T$.
This quantity is given by:
\begin{equation}
\left\langle j^{\nu }\right\rangle =e\,\mathrm{tr}[\hat{\rho }\bar{\psi}%
(x)\gamma ^{\nu }\psi (x)].  \label{C}
\end{equation}%
Substituting the expansion (\ref{psiexp}) and using the relations (\ref{traa}%
), similar to Eq. (\ref{FCexp}), the following mode-sum formula is obtained:%
\begin{equation}
\left\langle j^{\nu }\right\rangle =\left\langle j^{\nu }\right\rangle
_{0}+e\sum_{\sigma }\sum_{j=+,-}\frac{j\bar{\psi}_{\sigma }^{(j)}\gamma
^{\nu }\psi _{\sigma }^{(j)}}{e^{\beta (\varepsilon _{\sigma }^{(j)}-j\tilde{%
\mu })}+1},  \label{C1}
\end{equation}%
with%
\begin{equation}
\left\langle j^{\nu }\right\rangle _{0}=e\sum_{\beta }\bar{\psi}_{\beta
}^{(-)}(x)\gamma ^{\nu }\psi _{\beta }^{(-)}(x),  \label{Cvev}
\end{equation}%
being the corresponding vacuum expectation value. Details of the
calculations here are similar to those for the FC, and we will describe the
main steps only.

Firstly, we consider the charge density corresponding to the component $\nu
=0$. As it has been shown in Ref. \cite{Bell10}, the renormalized vacuum
expectation value for the charge density vanishes: $\left\langle
j^{0}\right\rangle _{0,\mathrm{ren}}=0$. By taking into account Eq. (\ref%
{psiplm}), for the finite temperature part of the charge density we get%
\begin{equation}
\left\langle j^{0}\right\rangle =\left\langle j^{0}\right\rangle
_{+}+\left\langle j^{0}\right\rangle _{-},  \label{C0dec}
\end{equation}%
where%
\begin{equation}
\left\langle j^{0}\right\rangle _{\pm }=\pm \frac{(4\pi )^{-p/2}eN}{\Gamma
(p/2)V_{q}}\sum_{\mathbf{n}_{q}\in \mathbf{Z}^{q}}\int_{\varepsilon _{%
\mathbf{n}_{q}}}^{\infty }dx\,\frac{x(x^{2}-\varepsilon _{\mathbf{n}%
_{q}}^{2})^{p/2-1}}{e^{\beta (x\mp \mu )}+1},  \label{C0}
\end{equation}%
are the contributions to the charge density from the particles (upper sign)
and antiparticles (lower sign). Note that for the number densities of
particles and antiparticles one has $\left\langle N\right\rangle _{\pm }=\pm
\left\langle j^{0}\right\rangle _{\pm }/e$. As it is seen, the signs of $%
\left\langle j^{0}\right\rangle $ and $\mu $ coincide, so $\left\langle
j^{0}\right\rangle \mu >0$.

In the case $|\mu |<$ $\varepsilon _{0}$, by using the expansion (\ref%
{Expansion}), the integration over $\mathbf{k}_{p}$ is explicitly done and
one finds%
\begin{equation}
\left\langle j^{0}\right\rangle _{\pm }=\mp \frac{eN\beta }{(2\pi
)^{(p+1)/2}V_{q}}\sum_{n=1}^{\infty }(-1)^{n}ne^{\pm n\beta \mu }\sum_{%
\mathbf{n}_{q}\in \mathbf{Z}^{q}}\varepsilon _{\mathbf{n}%
_{q}}^{p+1}f_{(p+1)/2}(n\beta \varepsilon _{\mathbf{n}_{q}}).  \label{C2pm}
\end{equation}%
For the total charge density we get%
\begin{equation}
\left\langle j^{0}\right\rangle =-\frac{2eN\beta }{(2\pi )^{(p+1)/2}V_{q}}%
\sum_{n=1}^{\infty }(-1)^{n}n\sinh (n\beta \mu )\sum_{\mathbf{n}_{q}\in
\mathbf{Z}^{q}}\varepsilon _{\mathbf{n}_{q}}^{p+1}f_{(p+1)/2}(n\beta
\varepsilon _{\mathbf{n}_{q}}).  \label{C2}
\end{equation}%
The charge density is an even periodic function of the phases $\tilde{\alpha
}_{l}$ with the period equal to the unity. Also, it is an odd function of
the chemical potential $\mu $. If the lengths of the compact dimensions are
large, the contribution of large $|n_{l}|$ dominates in Eq. (\ref{C2}) and,
to the leading order, we replace the summation over $\mathbf{n}_{q}$ by the
integration. By using Eq. (\ref{IntFormu}), we see that the leading term
coincides with the charge density in the topologically trivial Minkowski
spacetime:%
\begin{equation}
\left\langle j^{0}\right\rangle \approx \left\langle j^{0}\right\rangle _{%
\mathrm{M}}=-\frac{2eNm^{D+1}\beta }{(2\pi )^{(D+1)/2}}\sum_{n=1}^{\infty
}(-1)^{n}n\sinh (n\beta \mu )f_{(D+1)/2}(n\beta m).  \label{j0M}
\end{equation}

For small values of the length of the $l$th compact dimension, $L_{l}$, the
behavior of the charge density crucially depends on the value of the phase $%
\tilde{\alpha }_{l}$. For integer values of $\tilde{\alpha }_{l}$ the
dominant contribution to the sum over $n_{l}$ in Eq. (\ref{C2}) comes from
the term $n_{l}=-\tilde{\alpha }_{l}$ and, in the leading order, we obtain $%
\left\langle j^{0}\right\rangle \approx N\left\langle j^{0}\right\rangle
_{p,q-1}/(N_{D-1}L_{l})$, where $\left\langle j^{0}\right\rangle _{p,q-1}$
is the charge density in $(D-1)$-dimensional space with the spatial topology
$R^{p}\times (S^{1})^{q-1}$ and with the lengths of compact dimensions $%
(L_{p+1},\ldots ,L_{l-1},L_{l+1},\ldots ,L_{D})$. If $\tilde{\alpha }_{l}$
is not an integer, the dominant contribution to the sum over $n_{l}$ comes
from the term for which $|n_{l}+\tilde{\alpha }_{l}|$ is the smallest. In
this case, the argument of the function $f_{(p+1)/2}(x)$ in Eq. (\ref{C2})
is large and we can use the asymptotic expression for the Macdonald function
for large values of the argument. In this way we can see that the charge is
suppressed by the factor $e^{-\beta \bar{\alpha }_{l}/L_{l}}$, where $\bar{%
\alpha }_{l}=\mathrm{min}|n_{l}+\tilde{\alpha }_{l}|$. Finally, at low
temperatures, the dominant contribution in Eq. (\ref{C2}) comes from the
term with $n=1$ and from the term in the sum over $\mathbf{n}_{q}$ for which
$\varepsilon _{\mathbf{n}_{q}}$ is the smallest, $\varepsilon _{\mathbf{n}%
_{q}}=\varepsilon _{0}$. To the leading order we get%
\begin{equation}
\left\langle j^{0}\right\rangle =\frac{eN\,\mathrm{sgn}(\mu )\varepsilon
_{0}^{p/2}}{2(2\pi )^{p/2}V_{q}\beta ^{p/2}}e^{-\beta (\varepsilon _{0}-|\mu
|)},  \label{Clow}
\end{equation}%
and the charge is exponentially suppressed.

An equivalent expression for the charge density is obtained by using the
zeta function approach. In order to do that, first let us write the
expression (\ref{C0}) in the form%
\begin{equation}
\left\langle j^{0}\right\rangle =-\frac{eN}{V_{q}}\int \frac{d\mathbf{k}_{p}%
}{(2\pi )^{p}}\sum_{\mathbf{n}_{q}\in \mathbf{Z}^{q}}\sum_{n=1}^{\infty
}(-1)^{n}\sinh (n\beta \mu )e^{-n\beta \varepsilon (\mathbf{k})}.
\label{CZeta0}
\end{equation}%
Then it can be transformed as
\begin{equation}
\left\langle j^{0}\right\rangle =-\frac{eN}{V_{q}}\left( \mu -\int_{0}^{\mu
}d\mu \,\partial _{\beta }\beta \right) \int \frac{d\mathbf{k}_{p}}{(2\pi
)^{p}}\sum_{\mathbf{n}_{q}\in \mathbf{Z}^{q}}\sum_{n=1}^{\infty
}(-1)^{n}\cosh (n\beta \mu )\frac{e^{-n\beta \varepsilon (\mathbf{k})}}{%
\varepsilon (\mathbf{k})}.  \label{CZeta}
\end{equation}%
Now, by making use of Eq. (\ref{Sum1}), with the transformations similar to
those we have described for the case of the FC, the charge is presented as .%
\begin{equation}
\left\langle j^{0}\right\rangle =-eN\left( \mu -\int_{0}^{\mu }d\mu
\,\partial _{\beta }\beta \right) \zeta (1),  \label{CZ1}
\end{equation}%
with the zeta function defined by Eq. (\ref{Zeta}). Substituting the
expression (\ref{Zeta1}) for the zeta function, we see that the contribution
of the first term in the right-hand side of Eq. (\ref{Zeta1}) vanishes. The
contribution of the second term is finite at the physical point and for the
charge density we directly get%
\begin{eqnarray}
\left\langle j^{0}\right\rangle &=&-\frac{2eNm^{D+1}\beta }{(2\pi )^{(D+1)/2}%
}\sum_{n=1}^{\infty }(-1)^{n}n\sinh (n\beta \mu )\sum_{\mathbf{m}_{q}\in
\mathbf{Z}^{q}}\cos (2\pi \mathbf{m}_{q}\cdot \boldsymbol{\alpha }_{q})
\notag \\
&&\times f_{(D+1)/2}(m\sqrt{g^{2}(\mathbf{L}_{q},\mathbf{m}_{q})+n^{2}\beta
^{2}}).  \label{CZ2}
\end{eqnarray}%
Here, the term with $\mathbf{m}_{q}=0$ corresponds to the charge density in
topologically trivial Minkowski spacetime given by Eq. (\ref{j0M}). The
equivalence of the representations (\ref{C2}) and (\ref{CZ2}), can be seen
by using Eq. (\ref{FormulEquiv}).

An alternative representation, convenient for the investigation of the
high-temperature asymptotic of the charge density, is obtained from Eq. (\ref%
{CZ2}) with the help of formula (\ref{SumHighT}). Firstly, we separate from
the right-hand side of Eq. (\ref{CZ2}) the part corresponding to $%
\left\langle j^{0}\right\rangle _{\mathrm{M}}$. Then, by using $n\beta \sinh
(n\beta \mu )=\partial _{\mu }\cosh (n\beta \mu )$, we apply, for the sum
over $n$, the formula (\ref{SumHighT}). This yields:%
\begin{eqnarray}
\left\langle j^{0}\right\rangle &=&\left\langle j^{0}\right\rangle _{\mathrm{%
M}}-\frac{eNT}{(2\pi )^{D/2}}\partial _{\mu }\sum_{n=-\infty }^{+\infty }%
\left[ (\pi \left( 2n+1\right) T+i\mu )^{2}+m^{2}\right] ^{D/2}  \notag \\
&&\times \sideset{}{'}{\sum}_{\mathbf{m}_{q}\in \mathbf{Z}^{q}}\cos (2\pi
\mathbf{m}_{q}\cdot \boldsymbol{\alpha }_{q})f_{D/2}(g(\mathbf{L}_{q},%
\mathbf{m}_{q})\sqrt{(\pi \left( 2n+1\right) T+i\mu )^{2}+m^{2}}).
\label{CZ3}
\end{eqnarray}%
In the limit of high temperatures the contributions of the terms with $n=0$
and $n=-1$ dominate, and the part in the charge density induced by
nontrivial topology is suppressed by the factor $e^{-\pi L_{\mathrm{min}}T}$
and $\left\langle j^{0}\right\rangle \approx \left\langle j^{0}\right\rangle
_{\mathrm{M}}$. The high-temperature asymptotic of the Minkowskian part is
given by%
\begin{equation}
\left\langle j^{0}\right\rangle _{\mathrm{M}}\approx \frac{1-2^{2-D}}{\pi
^{(D+1)/2}}\Gamma ((D+1)/2)\zeta _{\mathrm{R}}(D-1)eN\mu T^{D-1}.
\label{j0MhighT}
\end{equation}%
For $D=2$ one has $\left\langle j^{0}\right\rangle _{\mathrm{M}}\approx
(2\pi )^{-1}eN\mu T\ln 2$.

Another representation for the charge density is obtained by applying to the
sum over $n_{l}$ in Eq. (\ref{CZeta0}) the summation formula (\ref{AP}) with
the functions $g(x)=1$ and $f(x)=\exp (-n\beta \sqrt{x^{2}+\mathbf{k}%
_{p}^{2}+\varepsilon _{\mathbf{n}_{q-1}^{l}}^{2}})$. After the
transformations similar to those for the FC, we get
\begin{eqnarray}
\left\langle j^{0}\right\rangle &=&\left\langle j^{0}\right\rangle
_{p+1,q-1}-\frac{4eN\beta L_{l}}{(2\pi )^{p/2+1}V_{q}}\sum_{n=1}^{\infty
}(-1)^{n}n\sinh (n\beta \mu )\sum_{r=1}^{\infty }\cos (2\pi r\tilde{\alpha }%
_{l})  \notag \\
&&\times \sum_{\mathbf{n}_{q-1}^{l}\in \mathbf{Z}^{q-1}}\varepsilon _{%
\mathbf{n}_{q-1}^{l}}^{p+2}f_{p/2+1}(\varepsilon _{\mathbf{n}_{q-1}^{l}}%
\sqrt{r^{2}L_{l}^{2}+n^{2}\beta ^{2}}),  \label{CAP}
\end{eqnarray}%
where $\left\langle j^{0}\right\rangle _{p+1,q-1}$ is the charge density in
the space with topology $R^{p+1}\times (S^{1})^{q-1}$ and with the lengths
of the compact dimensions $L_{p+1},\ldots ,L_{l-1},L_{l+1},\ldots ,L_{D}$.
The second term in the right-hand side of Eq. (\ref{CAP}) is induced by the
compactification of the coordinate $z^{l}$.

\begin{figure}[tbph]
\begin{center}
\begin{tabular}{cc}
\epsfig{figure=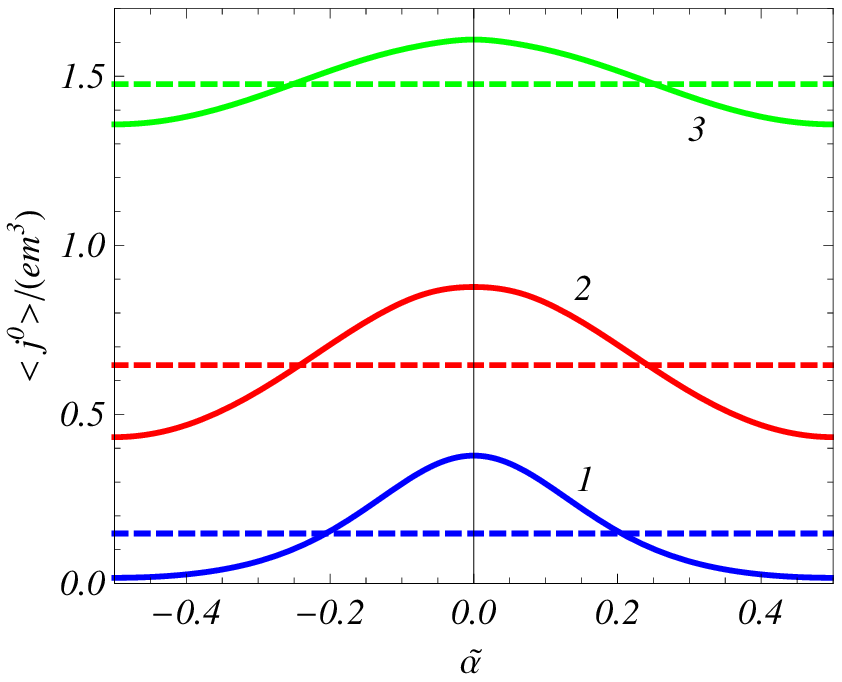,width=7.cm,height=6.cm} & \quad %
\epsfig{figure=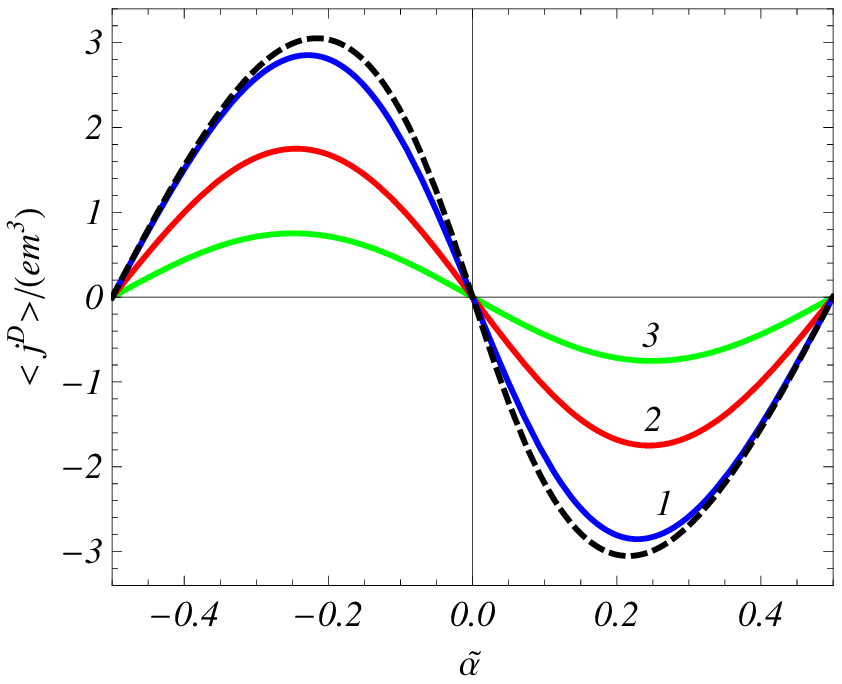,width=7.cm,height=6.cm}%
\end{tabular}%
\end{center}
\caption{Charge (left panel) and current (right panel) densities as
functions of $\tilde{\protect\alpha }$ in $D=3$ model with a single compact
dimension for the values of the parameters $\protect\mu /m=0.5 $, $%
mL_{3}=0.5 $. The curves are labelled with the value of $%
T/m$. On the left panel, the dashed lines correspond to the charge density
in Minkowski spacetime with trivial topology. The dashed curve on the right
panel presents the VEV of the current density.}
\label{fig1}
\end{figure}

In the left panel of figure \ref{fig1}, for the $D=3$ model with a single
compact dimension ($p=2$, $q=1$), we have plotted the charge density as a
function of $\tilde{\alpha }_{3}\equiv \tilde{\alpha }$ for the values of
the parameters $\mu /m=0.5$ and $mL_{3}=0.5$. The numbers near the curves
correspond to the values of $T/m$. The dashed lines present the charge
density in Minkowski spacetime with trivial topology.

For the case $|\mu |>$ $\varepsilon _{0}$, in a way similar to that we have
used for the FC, assuming that $\mu >0$, the charge density is presented as
\begin{eqnarray}
\left\langle j^{0}\right\rangle &=&\left\langle j^{0}\right\rangle _{-}+%
\frac{(4\pi )^{-p/2}eN}{\Gamma (p/2)V_{q}}\sum_{\varepsilon _{\mathbf{n}%
_{q}}\leqslant \varepsilon _{\mathbf{n}_{q}^{(0)}}}\int_{\varepsilon _{%
\mathbf{n}_{q}}}^{\infty }dx\,\frac{x(x^{2}-\varepsilon _{\mathbf{n}%
_{q}}^{2})^{p/2-1}}{e^{\beta (x-\mu )}+1}  \notag \\
&&-\frac{Ne\beta }{(2\pi )^{(p+1)/2}V_{q}}\sum_{n=1}^{\infty
}(-1)^{n}ne^{n\beta \mu }\sum_{\varepsilon _{\mathbf{n}_{q}}>\varepsilon _{%
\mathbf{n}_{q}^{(0)}}}\varepsilon _{\mathbf{n}_{q}}^{p+1}f_{(p+1)/2}(n\beta
\varepsilon _{\mathbf{n}_{q}}),  \label{C0mu}
\end{eqnarray}%
where the part coming from the antiparticles, $\left\langle
j^{0}\right\rangle _{-}$, is given by the same expression (\ref{C2pm}) as
before. In the limit $T\rightarrow 0$ we get%
\begin{equation}
\left\langle j^{0}\right\rangle =\frac{(4\pi )^{-p/2}eN}{p\Gamma (p/2)V_{q}}%
\sum_{\varepsilon _{\mathbf{n}_{q}}\leqslant \varepsilon _{\mathbf{n}%
_{q}^{(0)}}}(\mu ^{2}-\varepsilon _{\mathbf{n}_{q}}^{2})^{p/2}.  \label{j0T0}
\end{equation}%
Note that the number of states per unit volume with the energies smaller
than $\mu $ is equal to%
\begin{equation}
\mathcal{N}_{\leqslant \mu }=\frac{N}{2(2\pi )^{p}V_{q}}\sum_{\varepsilon _{%
\mathbf{n}_{q}}\leqslant \varepsilon _{\mathbf{n}_{q}^{(0)}}}\int_{|\mathbf{k%
}_{p}|\leqslant \sqrt{\mu ^{2}-\varepsilon _{\mathbf{n}_{q}}^{2}}}d\mathbf{k}%
_{p}.  \label{Nstates}
\end{equation}%
Now the charge density at zero temperatures can be written as $\left\langle
j^{0}\right\rangle _{T=0}=e\mathcal{N}_{\leqslant \mu }$. At high
temperatures, the leading term in the expansion of the charge density, as
before, is given by Eq. (\ref{j0MhighT}).

In the discussion above we have considered the charge density as a function
of the temperature, chemical potential, and the lengths of compact
dimensions. From the physical point of view it is also of interest to
consider the behavior of the system for a fixed value of the charge $%
Q=V_{q}\left\langle j^{0}\right\rangle $. In this case the expressions
derived in this section give the chemical potential as a function of the
charge, temperature and of the volume of the compact subspace. For example,
we can use the formula which follows from Eqs. (\ref{C0dec}) and (\ref{C0}):%
\begin{equation}
Q=\frac{(4\pi )^{-p/2}eN}{\Gamma (p/2)}\sum_{\mathbf{n}_{q}\in \mathbf{Z}%
^{q}}\int_{\varepsilon _{\mathbf{n}_{q}}}^{\infty }dx\,x\,\sum_{j=\pm }j%
\frac{(x^{2}-\varepsilon _{\mathbf{n}_{q}}^{2})^{p/2-1}}{e^{\beta (x-j\mu
)}+1}.  \label{Q}
\end{equation}%
From here it directly follows that the sign of $\mu $ coincides with the
sign of $Q$, consequently $\mu Q>0$, and that the chemical potential is an
odd function of $Q$. For fixed temperature and lengths of compact
dimensions, the function $|\mu |$ monotonically increases with increasing $%
|Q|$. At high temperatures, by taking into account that $Q\approx
V_{q}\left\langle j^{0}\right\rangle _{\mathrm{M}}$, up to exponentially
small terms, and using Eq. (\ref{j0MhighT}), in the leading order we get%
\begin{equation}
\mu \approx \frac{\left( 1-2^{2-D}\right) ^{-1}}{\Gamma ((D+1)/2)\zeta _{%
\mathrm{R}}(D-1)}\frac{\pi ^{(D+1)/2}Q}{eNV_{q}T^{D-1}}.  \label{muHightT}
\end{equation}%
With decreasing the temperature, the chemical potential increases and its
value at the zero temperature is determined from Eq. (\ref{j0T0}).

\begin{figure}[tbph]
\begin{center}
\begin{tabular}{cc}
\epsfig{figure=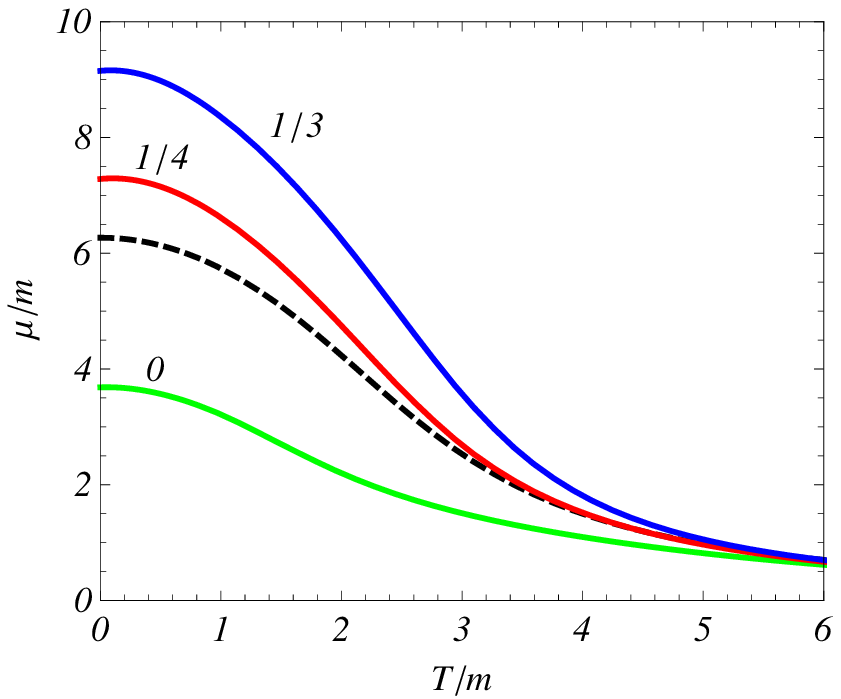,width=7.cm,height=6.cm} & \quad %
\epsfig{figure=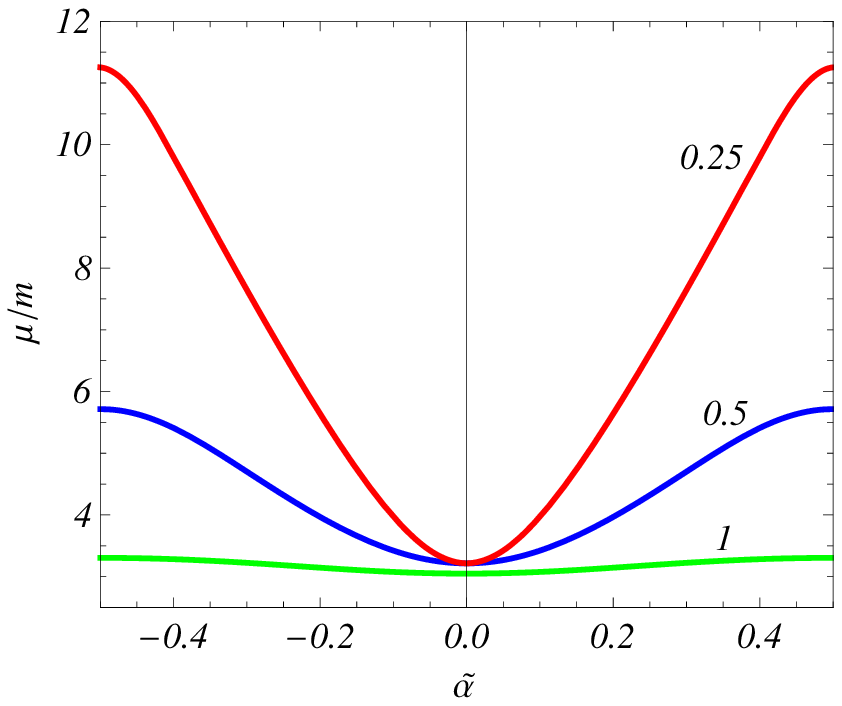,width=7.cm,height=6.cm}%
\end{tabular}%
\end{center}
\caption{The chemical potential as a function of the temperature (left
panel) and of the phase $\tilde{\protect\alpha}$ (right panel) in the $D=3$
model with a single compact dimension. The graphs are plotted for a fixed
value of the charge corresponding to $Q/(em^{2})=2$. For the left panel $%
mL_{3}=0.25$ and the numbers near the curves are the corresponding
values of the phase $\tilde{\protect\alpha}$. The dashed curve
presents the chemical potential in the Minkowski spacetime with
trivial topology. For the right panel $T/m=1$, and the curves are
labelled with the value of $mL_{3}$. } \label{fig2}
\end{figure}

On the left panel of figure \ref{fig2}, for the $D=3$ model with a single
compact dimension of the length $L_{3}$ ($p=2$, $q=1$), we plotted the
chemical potential as a function of the temperature for a fixed value of the
charge corresponding to $Q/(em^{2})=2$ and for $mL_{3}=0.25$. The numbers
near the curves correspond to the values of the phase $\tilde{\alpha}_{3}=%
\tilde{\alpha}$. The dashed curve presents the chemical potential in $D=3$
Minkowski spacetime with trivial topology $R^{3}$ and with the same charge
density corresponding to $\left\langle j^{0}\right\rangle _{\mathrm{M}%
}L_{3}/(em^{2})=2$. The right panel displays the chemical potential as a
function of $\tilde{\alpha}$ for the same value of the charge $Q/(em^{2})=2$
and for $T/m=1$. The numbers near the curves correspond to the values of $%
mL_{3}$.

\section{Current density}

\label{sec:Current}

Now let us consider the spatial components of the expectation value $%
\left\langle j^{\nu }\right\rangle $. Substituting the mode functions (\ref%
{psiplm}) into Eq. (\ref{C1}) we get%
\begin{equation}
\left\langle j^{\nu }\right\rangle =\left\langle j^{\nu }\right\rangle
_{0}+\sum_{j=+,-}\left\langle j^{\nu }\right\rangle _{j},  \label{jnudec}
\end{equation}%
where the contributions from the particles ($j=+$) and antiparticles ($j=-$)
are given by the expression below,
\begin{equation}
\left\langle j^{\nu }\right\rangle _{\pm }=\frac{eN}{2(2\pi )^{p}V_{q}}\int d%
\mathbf{k}_{p}\sum_{\mathbf{n}_{q}\in \mathbf{Z}^{q}}\frac{\tilde{k}_{\nu
}/\varepsilon (\mathbf{k})}{e^{\beta (\varepsilon (\mathbf{k})\mp \mu )}+1},
\label{jnu}
\end{equation}%
for $\nu =1,\ldots ,D$. In the case of $\nu =1,\ldots ,p$, the integrand is
an odd function of $k_{\nu }$ and the integral is zero. Hence, the current
density along uncompactified dimensions vanishes: $\left\langle j^{\nu
}\right\rangle =0$, $\nu =1,\ldots ,p$.

Under the condition $|\mu |<$ $\varepsilon _{0}$, by using Eq. (\ref%
{Expansion}), and after the integration over $\mathbf{k}_{p}$, we find%
\begin{equation}
\left\langle j^{\nu }\right\rangle _{\pm }=-\frac{eN}{(2\pi )^{(p+1)/2}V_{q}}%
\sum_{n=1}^{\infty }(-1)^{n}e^{\pm n\beta \mu }\sum_{\mathbf{n}_{q}\in
\mathbf{Z}^{q}}\tilde{k}_{\nu }\varepsilon _{\mathbf{n}%
_{q}}^{p-1}f_{(p-1)/2}(n\beta \varepsilon _{\mathbf{n}_{q}}),  \label{jnupm}
\end{equation}%
for the current densities coming from particles/antiparticles, and%
\begin{equation}
\left\langle j^{\nu }\right\rangle =\left\langle j^{\nu }\right\rangle _{0}-%
\frac{2eN}{(2\pi )^{(p+1)/2}V_{q}}\sum_{n=1}^{\infty }(-1)^{n}\cosh (n\beta
\mu )\sum_{\mathbf{n}_{q}\in \mathbf{Z}^{q}}\tilde{k}_{\nu }\varepsilon _{%
\mathbf{n}_{q}}^{p-1}f_{(p-1)/2}(n\beta \varepsilon _{\mathbf{n}_{q}}),
\label{jnu1}
\end{equation}%
for the total current density. From Eq. (\ref{jnu1}) it follows that the
current density along the $\nu $th compact dimension is an even function of
the phases $\tilde{\alpha }_{l}$, $l\neq \nu $, and an odd function of the
phase $\tilde{\alpha }_{\nu }$. In particular, $\left\langle j^{\nu
}\right\rangle =0$ for $\tilde{\alpha }_{\nu }=0$. The latter is the case
for twisted and untwisted fields in the absence of magnetic fluxes. At low
temperatures regime, the dominant contribution to the finite temperature
part comes from the mode with the lowest energy and, to the leading order,
one has%
\begin{equation}
\left\langle j^{\nu }\right\rangle =\left\langle j^{\nu }\right\rangle _{0}+%
\frac{eN\tilde{\alpha }_{\nu }\varepsilon _{0}^{p/2-1}e^{-\beta (\varepsilon
_{0}-|\mu |)}}{2(2\pi )^{p/2-1}V_{q}L_{\nu }\beta ^{p/2}}.  \label{jlowT}
\end{equation}%
In this limit, the finite temperature part is exponentially suppressed.

Another form for the current density is obtained on the base of the zeta
function approach. With this aim, we write Eq. (\ref{jnu}) as%
\begin{equation}
\left\langle j^{\nu }\right\rangle =-\frac{eN}{(2\pi )^{p}V_{q}}\int d%
\mathbf{k}_{p}\sum_{\mathbf{n}_{q}\in \mathbf{Z}^{q}}\frac{\tilde{k}_{\nu }}{%
\varepsilon (\mathbf{k})}\sideset{}{'}{\sum}_{n=0}^{\infty }(-1)^{n}\cosh
(n\beta \mu )e^{-n\beta \varepsilon (\mathbf{k})}.  \label{jnu2}
\end{equation}%
By using Eqs. (\ref{IntRep}) and (\ref{Sum1}), this expression is presented
in the form%
\begin{equation}
\left\langle j^{\nu }\right\rangle =-\frac{eN}{L_{\nu }}\sum_{n_{\nu
}=-\infty }^{+\infty }\tilde{k}_{\nu }\zeta _{\nu }(1),  \label{jnu3}
\end{equation}%
where the partial zeta function is defined as%
\begin{equation}
\zeta _{\nu }(s)=\frac{L_{\nu }}{\beta V_{q}}\int \frac{d\mathbf{k}_{p}}{%
(2\pi )^{p}}\sum_{\mathbf{n}_{q}^{\nu }\in \mathbf{Z}^{q}}(\mathbf{k}%
_{p}^{2}+\sum_{l=p+1}^{D+1}\tilde{k}_{l}^{2}+m^{2})^{-s},  \label{Zetanu}
\end{equation}%
with $\mathbf{n}_{q}^{\nu }=(n_{p+1},\ldots ,n_{\nu -1},n_{\nu +1},\ldots
,n_{D+1})$. Here, $\tilde{k}_{D+1}=2\pi (n_{D+1}+\tilde{\alpha }%
_{D+1})/L_{D+1}$ with $L_{D+1}$ and $\tilde{\alpha }_{D+1}$ being given by
Eq. (\ref{LD1}).

After the integration over $\mathbf{k}_{p}$ and the application of the
generalized Chowla-Selberg formula \cite{Eliz98}, the following
representation is obtained:%
\begin{eqnarray}
\zeta _{\nu }(s) &=&\frac{m_{\nu }^{D-2s}}{(4\pi )^{D/2}}\frac{\Gamma (s-D/2)%
}{\Gamma (s)}+\frac{2^{1-s}m_{\nu }^{D-2s}}{(2\pi )^{D/2}\Gamma (s)}  \notag
\\
&&\times \sideset{}{'}{\sum}_{\mathbf{n}_{q}^{\nu }\in \mathbf{Z}^{q}}\cos
(2\pi \mathbf{n}_{q}^{\nu }\cdot \boldsymbol{\alpha }_{q}^{\nu
})f_{D/2-s}(m_{\nu }g(\mathbf{L}_{q}^{\nu },\mathbf{n}_{q}^{\nu })),
\label{Zetanu1}
\end{eqnarray}%
where $\mathbf{L}_{q}=(L_{p+1},\ldots ,L_{\nu -1},L_{\nu +1},\ldots
L_{D},\beta )$, $\boldsymbol{\alpha }_{q}^{\nu }=(\tilde{\alpha}%
_{p+1},\ldots ,\tilde{\alpha}_{\nu -1},\tilde{\alpha}_{\nu +1},\ldots ,%
\tilde{\alpha}_{D+1})$ and
\begin{equation}
m_{\nu }=\sqrt{\tilde{k}_{\nu }^{2}+m^{2}}.  \label{mnu}
\end{equation}%
The contribution of the second term in the right-hand side of Eq. (\ref%
{Zetanu1}) to the current density is finite at the physical point $s=1$. The
analytical continuation of the first term is obtained by applying the
summation formula (\ref{AP}) to the series over $n_{\nu }$. In this way, we
get%
\begin{eqnarray}
&&\frac{\Gamma (s-D/2)}{(4\pi )^{D/2}\Gamma (s)}\sum_{n_{\nu }=-\infty
}^{+\infty }\frac{\tilde{k}_{\nu }}{m_{\nu }^{2s-D}}  \notag \\
&&\quad =2L_{\nu }\frac{(2m)^{(D+1-2s)/2+1}}{(4\pi )^{(D+1)/2}\Gamma (s)}%
\sum_{n=1}^{\infty }\sin (2\pi n\tilde{\alpha}_{\nu })\frac{%
K_{(D+1-2s)/2+1}(nL_{\nu }m)}{(nL_{\nu })^{(D+1-2s)/2}}.  \label{relZet1}
\end{eqnarray}

Combining Eqs. (\ref{Zetanu1}) and (\ref{relZet1}), for the current density
along the $\nu $th compact dimension we find
\begin{eqnarray}
\left\langle j^{\nu }\right\rangle &=&\left\langle j^{\nu }\right\rangle
_{D-1,1}-\frac{2eNL_{\nu }^{-1}}{(2\pi )^{D/2}}\sideset{}{'}{\sum}%
_{n=0}^{\infty }(-1)^{n}\cosh (n\beta \mu )\sum_{n_{\nu }=-\infty }^{+\infty
}\tilde{k}_{\nu }m_{\nu }^{D-2}  \notag \\
&&\times \sum_{\mathbf{n}_{q-1}^{\nu }\in \mathbf{Z}^{q-1}}\cos (2\pi
\mathbf{n}_{q-1}^{\nu }\cdot \boldsymbol{\alpha }_{q-1}^{\nu
})f_{D/2-1}(m_{\nu }(\sum_{i=p+1,\neq \nu }^{D}L_{i}^{2}n_{i}^{2}+n^{2}\beta
^{2})^{1/2}).  \label{jnu4}
\end{eqnarray}%
where the term $n=0$, $\mathbf{n}_{q-1}^{\nu }=0$ is excluded, $\mathbf{n}%
_{q-1}^{\nu }=(n_{p+1},\ldots ,n_{\nu -1},n_{\nu +1},\ldots ,n_{D})$, $%
\boldsymbol{\alpha }_{q-1}^{\nu }=(\tilde{\alpha}_{p+1},\ldots ,\tilde{\alpha%
}_{\nu -1},\tilde{\alpha}_{\nu +1},\ldots \tilde{\alpha}_{D})$, and
\begin{equation}
\left\langle j^{\nu }\right\rangle _{D-1,1}=-\frac{2eNm^{D+1}L_{\nu }}{(2\pi
)^{(D+1)/2}}\sum_{l=1}^{\infty }l\sin (2\pi l\tilde{\alpha}_{\nu
})f_{(D+1)/2}(lL_{\nu }m),  \label{jnu1d}
\end{equation}%
is the zero temperature current density for the topology $R^{D-1}\times
S^{1} $ with the compact dimension of the length $L_{\nu }$. A more compact
form is obtained by making use of the relation%
\begin{eqnarray}
&&\sum_{n_{\nu }=-\infty }^{+\infty }\tilde{k}_{\nu }m_{\nu
}^{D-2}f_{D/2-1}(m_{\nu }y)  \notag \\
&&\quad =\sqrt{\frac{2}{\pi }}L_{\nu }^{2}m^{D+1}\sum_{n_{\nu }=1}^{\infty
}n_{\nu }\sin (2\pi n_{\nu }\tilde{\alpha}_{\nu })f_{(D+1)/2}(m\sqrt{%
y^{2}+n_{\nu }^{2}L_{\nu }^{2}}).  \label{Rel3}
\end{eqnarray}%
This leads to the final result:%
\begin{eqnarray}
\left\langle j^{\nu }\right\rangle &=&-\frac{4eNm^{D+1}L_{\nu }}{(2\pi
)^{(D+1)/2}}\sideset{}{'}{\sum}_{n=0}^{\infty }(-1)^{n}\cosh (n\beta \mu
)\sum_{n_{\nu }=1}^{\infty }n_{\nu }\sin (2\pi n_{\nu }\tilde{\alpha}_{\nu })
\notag \\
&&\times \sum_{\mathbf{n}_{q-1}^{\nu }\in \mathbf{Z}^{q-1}}\cos (2\pi
\mathbf{n}_{q-1}^{\nu }\cdot \boldsymbol{\alpha }_{q-1}^{\nu })f_{(D+1)/2}(m%
\sqrt{g^{2}(\mathbf{L}_{q},\mathbf{n}_{q})+n^{2}\beta ^{2}}).  \label{jnu6}
\end{eqnarray}%
The $n=0$ term in this formula corresponds to the VEV of the current
density:
\begin{eqnarray}
\left\langle j^{\nu }\right\rangle _{0} &=&-\frac{2eNm^{D+1}L_{\nu }}{(2\pi
)^{(D+1)/2}}\sum_{n_{\nu }=1}^{\infty }n_{\nu }\sin (2\pi n_{\nu }\tilde{%
\alpha}_{\nu })  \notag \\
&&\times \sum_{\mathbf{n}_{q-1}^{\nu }\in \mathbf{Z}^{q-1}}\cos (2\pi
\mathbf{n}_{q-1}^{\nu }\cdot \boldsymbol{\alpha }_{q-1}^{\nu
})f_{(D+1)/2}(mg(\mathbf{L}_{q},\mathbf{n}_{q})).  \label{jnT0}
\end{eqnarray}

In order to see the asymptotic behavior of the current density at high
temperatures, we apply to the series over $n$ in Eq. (\ref{jnu6}) the
formula (\ref{SumHighT}). This leads to the expression
\begin{eqnarray}
\left\langle j^{\nu }\right\rangle &=&-\frac{2eNL_{\nu }T}{(2\pi )^{D/2}}%
\sum_{n_{\nu }=1}^{\infty }n_{\nu }\sin (2\pi n_{\nu }\tilde{\alpha}_{\nu
})\sum_{n=-\infty }^{+\infty }\left[ (\pi (2n+1)T+i\mu )^{2}+m^{2}\right]
^{D/2}  \notag \\
&&\times \sum_{\mathbf{n}_{q-1}^{\nu }\in \mathbf{Z}^{q-1}}\cos (2\pi
\mathbf{n}_{q-1}^{\nu }\cdot \boldsymbol{\alpha }_{q-1}^{\nu })f_{D/2}(g(%
\mathbf{L}_{q},\mathbf{n}_{q})\sqrt{(\pi (2n+1)T+i\mu )^{2}+m^{2}}).
\label{jnu7}
\end{eqnarray}%
At high temperatures the dominant contribution comes from the terms $n=-1,0$
in the series over $n$ and from the term with $n_{l}=\delta _{l\nu }$. To
the leading order, we get%
\begin{equation}
\left\langle j^{\nu }\right\rangle \approx -\frac{2eN\sin (2\pi \tilde{\alpha%
}_{\nu })T^{(D+1)/2}}{(2L_{\nu })^{(D-1)/2}e^{\pi L_{\nu }T}},
\label{jnuHihgT}
\end{equation}%
and the current density is exponentially suppressed. This is in sharp
contrast with the high-temperature behavior of the current density in the
case of a scalar field. At high temperatures, the current density for a
scalar field linearly grows with the temperature \cite{Beze13}. This
difference of the asymptotics for scalar and fermionic current densities is
a direct consequence of different periodicity conditions imposed on the
fields along imaginary time (periodic and antiperiodic conditions for scalar
and fermion fields, respectively).

An alternative representation for the current density is obtained by
applying the summation formula (\ref{AP}) to the series over $n_{\nu }$ in
Eq. (\ref{jnu2}) by taking $g(x)=x$ and the function $f(x)$ from Eq. (\ref%
{gf}). With this choice, the first term in the right-hand side of Eq. (\ref%
{AP}) vanishes, and for the current density one gets%
\begin{eqnarray}
\left\langle j^{\nu }\right\rangle &=&-\frac{4eNL_{\nu }^{2}}{(2\pi
)^{p/2+1}V_{q}}\sideset{}{'}{\sum}_{n=0}^{\infty }(-1)^{n}\cosh (n\beta \mu
)\sum_{l=1}^{\infty }l\sin (2\pi l\tilde{\alpha}_{\nu })  \notag \\
&&\times \sum_{\mathbf{n}_{q-1}^{\nu }\in \mathbf{Z}^{q-1}}\varepsilon _{%
\mathbf{n}_{q-1}^{\nu }}^{p+2}f_{p/2+1}(\varepsilon _{\mathbf{n}_{q-1}^{\nu
}}\sqrt{l^{2}L_{\nu }^{2}+n^{2}\beta ^{2}}).  \label{jnu5}
\end{eqnarray}%
The $n=0$ term in this expression corresponds to the VEV of the current
density:%
\begin{equation}
\left\langle j^{\nu }\right\rangle _{0}=-\frac{2eNL_{\nu }^{2}}{(2\pi
)^{p/2+1}V_{q}}\sum_{l=1}^{\infty }l\sin (2\pi l\tilde{\alpha}_{\nu })\sum_{%
\mathbf{n}_{q-1}^{\nu }\in \mathbf{Z}^{q-1}}\varepsilon _{\mathbf{n}%
_{q-1}^{\nu }}^{p+2}f_{p/2+1}(lL_{\nu }\varepsilon _{\mathbf{n}_{q-1}^{\nu
}}).  \label{jnu0}
\end{equation}%
In the model with a single compact dimension Eq. (\ref{jnu5}) is reduced to
Eq. (\ref{jnu1d}).

Let us consider the limit of Eq. (\ref{jnu5}) when the length of the $r$th
dimension is much smaller than $L_{\nu }$, $L_{r}\ll L_{\nu }$. The current
density is a periodic function of $\tilde{\alpha}_{r}$ with the period equal
to unity and, without loss of generality, we can assume that $|\tilde{\alpha}%
_{r}|<1/2$. If $\tilde{\alpha}_{r}=0$, the dominant contribution comes from
the term $n_{r}=0$ and, to the leading order, we have $\left\langle j^{\nu
}\right\rangle \approx N\left\langle j^{\nu }\right\rangle
_{(p,q-1)}/(N_{D-1}L_{r})$, where $\left\langle j^{\nu }\right\rangle
_{(p,q-1)}$ is the current density in $(D-1)$-dimensional space of topology $%
R^{p}\times (S^{1})^{q-1}$ with the lengths of the compact dimensions $%
L_{p+1}$,\ldots ,$L_{r-1}$,$L_{r+1}$,\ldots ,$L_{D}$. The corrections to
this leading term are of the order $e^{-2\pi L_{\nu }/L_{r}}$. For $\tilde{%
\alpha}_{r}\neq 0$, once again, the dominant contribution comes from the
term with $n_{r}=0$, and the argument of the function $f_{p/2+1}(x)$ in Eq. (%
\ref{jnu5}) is large. By using the corresponding asymptotic for the
Macdonald function, we see that the current density is suppressed by the
factor $e^{-2\pi |\tilde{\alpha}_{r}|L_{\nu }/L_{r}}$.

The right panel of figure \ref{fig1}, displays the current density in the $%
D=3$ model with a single compact dimension, as a function of $\tilde{\alpha}%
_{3}\equiv \tilde{\alpha}$ for $\mu /m=0.5$ and $mL_{3}=0.5$. The numbers
near the curves are the correspond values of $T/m$. The dashed curve
presents the current density at zero temperature.

Now we turn to the case $|\mu |>$ $\varepsilon _{0}$. After the integration
in Eq. (\ref{jnu}), the general formula (\ref{jnudec}) is given in the form%
\begin{equation}
\left\langle j^{\nu }\right\rangle =\left\langle j^{\nu }\right\rangle _{0}+%
\frac{(4\pi )^{-p/2}eN}{\Gamma (p/2)V_{q}}\sum_{\mathbf{n}_{q}\in \mathbf{Z}%
^{q}}\tilde{k}_{\nu }\int_{\varepsilon _{\mathbf{n}_{q}}}^{\infty
}dx\,\sum_{j=+,-}\frac{(x^{2}-\varepsilon _{\mathbf{n}_{q}}^{2})^{p/2-1}}{%
e^{\beta (x-j\mu )}+1}.  \label{jnun}
\end{equation}%
For definiteness, assuming that $\mu >0$, the transformations for the part
with $j=-$ and for the part $j=+$ in the range $\varepsilon _{\mathbf{n}%
_{q}}>\mu $ are the same as those described above. For the current density
along the $\nu $th compact dimension one finds:
\begin{eqnarray}
\left\langle j^{\nu }\right\rangle &=&\left\langle j^{\nu }\right\rangle
_{0}+\left\langle j^{\nu }\right\rangle _{-}+\frac{(4\pi )^{-p/2}eN}{\Gamma
(p/2)V_{q}}\sum_{\varepsilon _{\mathbf{n}_{q}}\leqslant \varepsilon _{%
\mathbf{n}_{q}^{(0)}}}\tilde{k}_{\nu }\int_{\varepsilon _{\mathbf{n}%
_{q}}}^{\infty }dx\,\frac{(x^{2}-\varepsilon _{\mathbf{n}_{q}}^{2})^{p/2-1}}{%
e^{\beta (x-\mu )}+1}  \notag \\
&&-\frac{eN}{(2\pi )^{(p+1)/2}V_{q}}\sum_{n=1}^{\infty }(-1)^{n}e^{n\beta
\mu }\sum_{\varepsilon _{\mathbf{n}_{q}}>\varepsilon _{\mathbf{n}_{q}^{(0)}}}%
\tilde{k}_{\nu }\varepsilon _{\mathbf{n}_{q}}^{p-1}f_{(p-1)/2}(n\beta
\varepsilon _{\mathbf{n}_{q}}).  \label{jnimu}
\end{eqnarray}%
In the limit $T\rightarrow 0$ we get%
\begin{equation}
\left\langle j^{\nu }\right\rangle _{T=0}=\left\langle j^{\nu }\right\rangle
_{0}+\frac{(4\pi )^{-p/2}eN}{V_{q}\Gamma (p/2)}\sum_{\varepsilon _{\mathbf{n}%
_{q}}\leqslant \varepsilon _{\mathbf{n}_{q}^{(0)}}}\tilde{k}_{\nu
}\int_{\varepsilon _{\mathbf{n}_{q}}}^{\mu }dx\,(x^{2}-\varepsilon _{\mathbf{%
n}_{q}}^{2})^{p/2-1}.  \label{jnumuT0}
\end{equation}%
The second term on the right is the current density induced by the particles
filling the states with the energies $\varepsilon _{\mathbf{n}_{q}}\leq \mu $%
. At zero temperature the states with the energies $\varepsilon _{\mathbf{n}%
_{q}}>\mu $ are empty. At high temperatures, similar to the case $|\mu |<$ $%
\varepsilon _{0}$, the current density is exponentially suppressed (see Eq. (%
\ref{jnuHihgT})).

\begin{figure}[tbph]
\begin{center}
\begin{tabular}{cc}
\epsfig{figure=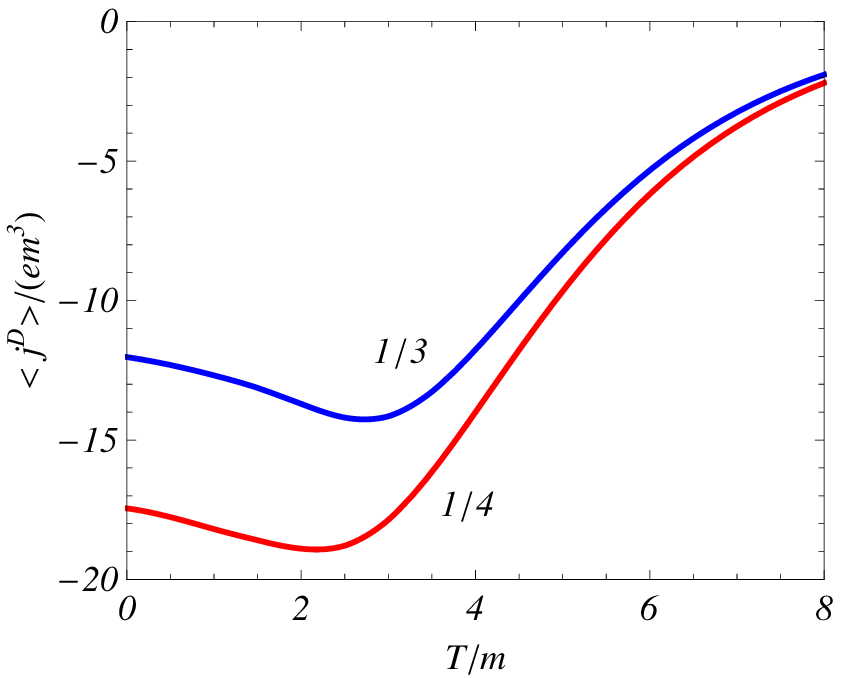,width=7.cm,height=6.cm} & \quad %
\epsfig{figure=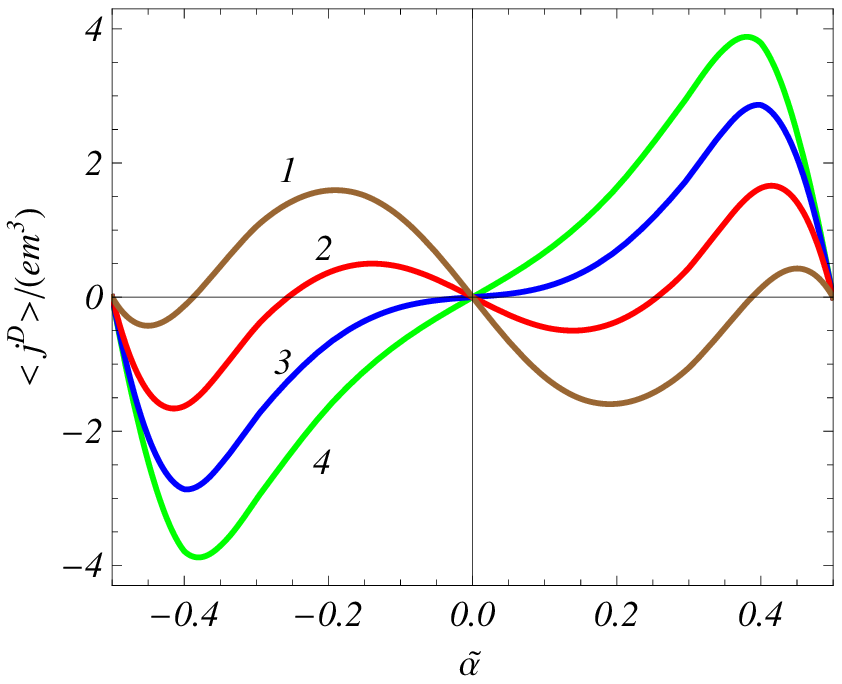,width=7.cm,height=6.cm}%
\end{tabular}%
\end{center}
\caption{The current density for a fixed value of the charge as a
function of the temperature (left panel) and of the phase
$\tilde{\protect\alpha}$ (right panel) in $D=3$ model with a
single compact dimension. On the left panel the graphs are plotted
for $Q/(em^{2})=2$, $mL_{3}=0.25$ and the curves are labelled with
the value of $\tilde{\protect\alpha}$. For the right panel
$T/m=0.5$, $mL_{3}=0.5$, and the numbers near the curves
correspond to the values of $Q/(em^{2})$. } \label{fig3}
\end{figure}

For a fixed value of the charge, Eq. (\ref{Q}) determines the chemical
potential as a function of temperature. Examples of this function are
plotted in figure \ref{fig2}. Substituting the chemical potential into Eq. (%
\ref{jnun}) or Eq. (\ref{jnimu}), we find the current density as a function
of the temperature for a fixed value of the charge. On the left panel of
figure \ref{fig3} we have plotted this function in the $D=3$ model with a
single compact dimension of the length $L_{3}$, for the fixed value of the
charge corresponding to $Q/(em^{2})=2$. The graphs are plotted for $%
mL_{3}=0.25$ and the numbers near the curves correspond to the values of $%
\tilde{\alpha}_{3}=\tilde{\alpha }$. The right panel presents the current
density versus the phase $\tilde{\alpha}$ for different values of $Q/(em^{2})
$ (numbers near the curves) and for $T/m=0.5$, $mL_{3}=0.5$.

\section{Expectation values in time-reversal symmetric odd dimensional models%
}

\label{sec:OddDim}

In the discussion above we have considered a fermionic field realizing the
irreducible representation of the Clifford algebra. With this
representation, in odd dimensional spacetimes ($D$ is an even number) the
mass term in the Lagrangian breaks $C$-invariance in $D=4n$, $P$-invariance
in $D=4n,4n+2$, and $T$-invariance in $D=4n+2$ (for a general discussion see
Ref. \cite{Shim85}). In order to restore these symmetries, we note that in
odd dimensions the $\gamma ^{D}$ matrix can be represented by other gamma
matrices in the following way,%
\begin{equation}
\gamma ^{D}=\gamma _{\pm }^{D}=\left\{
\begin{array}{cc}
\pm \gamma , & D=4n, \\
\pm i\gamma , & D=4n+2,%
\end{array}%
\right.  \label{gamD}
\end{equation}%
where $\gamma =\gamma ^{0}\gamma ^{1}\cdots \gamma ^{D-1}$. Hence, the
Clifford algebra in odd dimensions has two inequivalent representations
corresponding to the upper and lower signs in Eq. (\ref{gamD}). Now, two $N$%
-component Dirac fields, $\psi _{+}$ and $\psi _{-}$, can be introduced with
the equations%
\begin{equation}
(i\gamma _{\pm }^{\mu }D_{\mu }-m)\psi _{\pm }=0,  \label{DirEqOdd}
\end{equation}%
where $\gamma _{\pm }^{\mu }=(\gamma ^{0},\gamma ^{1},\cdots \gamma
^{D-1},\gamma _{\pm }^{D})$. Defining the charge conjugation as $\psi
_{+}^{C}=C\bar{\psi}_{-}^{T}$ (for $C$, $P$ and $T$ transformations, see,
for instance, Ref. \cite{Shim85}), we see that the total Lagrangian $%
\mathcal{L}=\sum_{j=+,-}\bar{\psi}_{j}(i\gamma _{j}^{\mu }D_{\mu }-m)\psi
_{j}$ is invariant under the charge conjugation. In a similar way, by
suitable transformations of the fields it can be seen that the combined
Lagrangian is invariant under $P$ and $T$ transformations, as well (in the
absence of magnetic fields). Note that by defining new fields $\psi
_{+}^{\prime }=\psi _{+}$, $\psi _{-}^{\prime }=\gamma \psi _{-}$, the
Lagrangian is transformed to the form%
\begin{equation}
\mathcal{L}=\sum_{j=+,-}\bar{\psi}_{j}^{\prime }(i\gamma ^{\mu }D_{\mu
}-jm)\psi _{j}^{\prime }.  \label{Lodd}
\end{equation}%
Thus, the field $\psi _{-}^{\prime }$ satisfies the same equation as $\psi
_{+}^{\prime }$ with the opposite sign for the mass term. The $N$-component
Dirac spinors $\psi _{+}^{\prime }$ and $\psi _{-}^{\prime }$ can be
combined in a $2N$-component spinor: $\Psi =(\psi _{+}^{\prime },\psi
_{-}^{\prime })^{T}$. Introducing $2N\times 2N$ Dirac matrices $\tilde{\gamma%
}^{\mu }=I_{2}\otimes \gamma ^{\mu }$ and $\eta =\sigma _{\mathrm{P}%
3}\otimes I_{N}$, where $I_{N}$ is the $N\times N$ unit matrix and $\sigma _{%
\mathrm{P}3}=\mathrm{diag}(1,-1)$ the Pauli matrix, the Lagrangian is
written in the form $\mathcal{L}=\bar{\Psi }(i\tilde{\gamma}^{\mu }D_{\mu
}-m\eta )\Psi $. An alternative representation is obtained by using the $%
2N\times 2N$ reducible representation of gamma matrices $\gamma _{(2N)}^{\mu
}=\sigma _{3}\otimes \gamma ^{\mu }$ with the Lagrangian%
\begin{equation}
\mathcal{L}=\bar{\Psi }(i\gamma _{(2N)}^{\mu }D_{\mu }-m)\Psi .
\label{Loddb}
\end{equation}

The FC, the charge and current densities for the model described by the
Lagrangian (\ref{Lodd}) can be obtained from the formulas given above. In
deriving the expectation values of these quantities we have assumed that the
parameter $m$ is nonnegative. However, the results are easily generalized
for a negative $m$ as well. It can be seen that $\left\langle \bar{\psi}\psi
\right\rangle /m$ and $\left\langle j^{\mu }\right\rangle $ are even
functions of $m$. Consequently, the current densities corresponding to the
fields $\psi _{+}^{\prime }$ and $\psi _{-}^{\prime }$ are given by
expressions presented in Sections \ref{sec:Charge} and \ref{sec:Current}
whereas the FC for the fields $\psi _{+}^{\prime }$ and $\psi _{-}^{\prime }$
differ by the sign. Hence, assuming that in Eq. (\ref{Lodd}) $m\geqslant 0$,
for the total expectation values one finds%
\begin{eqnarray}
\left\langle \bar{\Psi }\Psi \right\rangle &=&\left\langle \bar{\psi}\psi
\right\rangle _{\uparrow }-\left\langle \bar{\psi}\psi \right\rangle
_{\downarrow },  \notag \\
\left\langle j^{\nu }\right\rangle &=&\left\langle j^{\nu }\right\rangle
_{\uparrow }+\left\langle j^{\nu }\right\rangle _{\downarrow },  \label{EVs}
\end{eqnarray}%
where the parts with the indices $\uparrow $ and $\downarrow $ are the
contributions coming from the upper ($\psi _{+}^{\prime }$) and lower ($\psi
_{-}^{\prime }$) components of the $2N$-component spinor $\Psi $. The
separate terms in (\ref{EVs}), $\left\langle \bar{\psi}\psi \right\rangle
_{\uparrow ,\downarrow }$ and $\left\langle j^{\nu }\right\rangle _{\uparrow
,\downarrow }$ are given by the expressions presented in Sections \ref%
{sec:Charge} and \ref{sec:Current}. If the phases in the quasiperiodicity
conditions for the upper and lower components of the $2N$-spinor $\Psi $ are
the same, then the resulting FC vanishes, whereas the expressions for the
charge and current densities are obtained from the expressions given above
with an additional coefficient 2. However, the phases in the
quasiperiodicity conditions for the upper and lower components, in general,
can be different. As we will see below, this is the case for the Dirac model
in a class of carbon nanotubes.

\section{Applications to carbon nanotubes}

\label{sec:Nano}

In this section we apply the general results given above for the
investigation of the FC and current density in carbon nanotubes (for a
review see Ref. \cite{Sait98}). A single-wall cylindrical nanotube is
obtained from a graphene sheet by rolling it into a cylindrical shape. In
graphene, the low-energy excitations of the electronic subsystem, for a
given value of spin $s=\pm 1$, are described by an effective Dirac theory of
4-component spinors $\Psi _{s}=(\psi _{+,As},\psi _{+,Bs},\psi _{-,As},\psi
_{-,Bs})$, where the components $\psi _{\pm ,Js}$, with $J=A$ and $J=B$,
give the amplitude of the wave function on the $A$ and $B$ triangular
sublattices of the graphene hexagonal lattice and the indices $+$ and $-$
correspond to inequivalent points $(\mathbf{K}_{+},\mathbf{K}_{-})$ at the
corners of the two-dimensional Brillouin zone (see Ref. \cite{Gusy07}). The
quasiparticles are confined to a graphene sheet and, hence, for the spatial
dimension of the corresponding Dirac-like theory one has $D=2$. The
corresponding Lagrangian has the form (we restore the standard units)
\begin{equation}
\mathcal{L}=\sum_{s=\pm 1}\bar{\Psi}_{s}(i\hbar \gamma ^{0}\partial
_{t}+i\hbar v_{F}\gamma ^{l}D_{l}-\Delta )\Psi _{s},  \label{Lgraph}
\end{equation}%
where $D_{l}=(\mathbf{\nabla }-ie\mathbf{A}/\hbar c)_{l}$, $l=1,2$, with $%
e=-|e|$ for electrons, and $v_{F}\approx 7.9\times 10^{7}$ cm/s is the Fermi
velocity which plays the role of the speed of light. (For other planar
condensed-matter systems with the low-energy sector described by Eq. (\ref%
{Lgraph}) see, for instance, Ref. \cite{Shar04}.) For the Fermi velocity one
has $v_{F}=\sqrt{3}a\gamma _{0}/(2\hbar )$, where $a$ is the lattice
constant and $\gamma _{0}\approx 2.9$ eV is the transfer integral between
first-neighbor $\pi $ orbitals. We have included in the Lagrangian the
energy gap $\Delta $. It is expressed in terms of the corresponding Dirac
mass $m_{D}$ as $\Delta =m_{D}v_{F}^{2}$. The gap in the energy spectrum is
essential in many physical applications and can be generated by a number of
mechanisms (see, for example, Ref. \cite{Gusy07} and references therein).
Some mechanisms give rise to mass terms with the matrix structure different
from that we consider here. In dependence of the physical mechanism for the
generation, the energy gap may vary in the range $1\,\mathrm{meV}\lesssim
\Delta \lesssim 1\,\mathrm{eV}$. In the discussion of the graphene
properties within the framework of the model described by Lagrangian (\ref%
{Lgraph}), the Dirac matrices are usually taken in the form $\gamma ^{\mu
}=\sigma _{\mathrm{P}3}\otimes (\sigma _{\mathrm{P}3},i\sigma _{\mathrm{P}%
2},-i\sigma _{\mathrm{P}1})$. Now comparing with Eq. (\ref{Loddb}), we see
that Eq. (\ref{Lgraph}) is a special case with $D=2$, $N=2$.

In the case of the cylindrical nanotube, the spatial topology for the
effective Dirac theory is $R^{1}\times S^{1}$ with the compactified
dimension of the length $L$. The carbon nanotube is characterized by its
chiral vector $\mathbf{C}_{h}=n_{c}\mathbf{a}_{1}+m_{c}\mathbf{a}_{2}$,
where $\mathbf{a}_{1}$ and $\mathbf{a}_{2}$ are the basis vectors of the
hexagonal lattice of graphene and $n_{c}$, $m_{c}$ are integers. The length
of the compact dimension is given by $L=|\mathbf{C}_{h}|=aN_{c}$, with $%
N_{c}=\sqrt{n_{c}^{2}+m_{c}^{2}+n_{c}m_{c}}$ and $a=|\mathbf{a}_{1}|=|%
\mathbf{a}_{2}|=2.46\mathring{A}$ being the lattice constant. The special
cases $\mathbf{C}_{h}=(n_{c},0)$ and $\mathbf{C}_{h}=(n_{c},n_{c})$
correspond to zigzag and armchair nanotubes respectively. All other cases
correspond to chiral nanotubes. In the case $n_{c}-m_{c}=3l_{c}$, $l_{c}\in Z
$, the nanotube will be metallic and in the case $n_{c}-m_{c}\neq 3l_{c}$
the nanotube will be a semiconductor with an energy gap inversely
proportional to the diameter. In particular, the armchair nanotube is
metallic, and the $(n_{c},0)$ zigzag nanotube is metallic if and only if $%
n_{c}$ is an integer multiple of 3.

Periodicity conditions along the compact dimension for the bi-spinor $\Psi
_{s}$ in (\ref{Lgraph}) are obtained from the periodicity of the electron
wave function (see Refs. \cite{Sait98,Ando05}). For metallic nanotubes one
has periodic boundary conditions ($\alpha _{l}=0$ in Eq. (\ref{PerCond}))
and for semiconducting nanotubes, depending on the chiral vector, we have
two classes of inequivalent boundary conditions corresponding to $\alpha
_{l}=\pm 1/3$ ($n_{c}-m_{c}=3l_{c}\pm 1$). The phases have opposite signs
for the upper and lower components of the 4-spinor $\Psi _{s}$ in (\ref%
{Lgraph}), corresponding to the points $\mathbf{K}_{+}$ and $\mathbf{K}_{-}$.

For a given value of the spin $s$, the expressions of the FC and of the
expectation values of the current densities for separate contributions from
the points $\mathbf{K}_{+}$ and $\mathbf{K}_{-}$ are obtained from the
expressions given in previous sections by the replacements%
\begin{equation}
m\rightarrow a^{-1}\Delta /\gamma _{F},\;\beta \rightarrow a\gamma _{F}\beta
,\;\mu \rightarrow a^{-1}\mu /\gamma _{F},  \label{Replace}
\end{equation}%
where $\gamma _{F}=\hbar v_{F}/a=\sqrt{3}\gamma _{0}/2\approx 2.51$ eV
determines the characteristic energy scale of the model. In addition, in
expressions for the current density along compact dimensions a factor $v_{F}$
should be added, because now the operator of the spatial components of the
current density is defined as $j^{\nu }=ev_{F}\bar{\psi}(x)\gamma ^{\nu
}\psi (x)$. For a given spin $s$, the separate contributions are combined in
a way given by Eq. (\ref{EVs}), where $\uparrow $ and $\downarrow $
correspond to the points $\mathbf{K}_{+}$ and $\mathbf{K}_{-}$,
respectively. In the model under consideration, the spin $s=+1$ and $s=-1$
give the same contributions to the total expectation values. For cylindrical
carbon nanotubes, the spatial topology for the effective theory corresponds
to $R^{1}\times S^{1}$ and, hence, in the formulas above $p=1$, $q=1$.

The compactification of the direction along the cylinder axis gives another
class of graphene made structures called toroidal carbon nanotubes \cite%
{Liu97} with the background topology $(S^{1})^{2}$ ($p=0,q=2$). The
nanotorus is classified by its chiral vector $\mathbf{C}_{h}$ and
translational vector $\mathbf{T}=p_{c}\mathbf{a}_{1}+q_{c}\mathbf{a}_{2}$
with the coordinates along these directions $z^{1}$ and $z^{2}$ and with the
lengths $L_{1}$ and $L_{2}$. Usually one has $L_{1}\ll L_{2}$. From the
asymptotic analysis given above (see the paragraph after Eq. (\ref{jnu0})),
it follows that for $\tilde{\alpha}_{1}=0$ the current density along $z^{2}$%
, in the leading order, coincides with the corresponding quantity in the
model with $D=1$ and with the length of the compact dimension $L_{2}$. The
corresponding corrections are of the order $e^{-2\pi L_{2}/L_{1}}$. In the
absence of the magnetic flux, this case corresponds to nanotubes for which $%
\alpha _{1}=0$ and, hence, $n_{c}-m_{c}=3l_{c}$, $l_{c}\in Z$, (type I and
II toroidal carbon nanotubes). For $\tilde{\alpha}_{1}\neq 0$, the current
density along $z^{2}$-direction is suppressed by the factor $e^{-2\pi |%
\tilde{\alpha}_{1}|L_{2}/L_{1}}$. In the absence of the magnetic flux, this
case corresponds to $n_{c}-m_{c}\neq 3l_{c}$, $l_{c}\in Z$, with $|\alpha
_{1}|=1/3$ (type III toroidal carbon nanotubes).

In discussing the FC and current density in cylindrical and toroidal
nanotubes we work within the zone-folding approximation in which the effect
of confinement is to induce the selection on allowed values of the momentum
projection along the compact dimension. This approximation ignores the
curvature effects which may cause the mixing of $\sigma $ and $\pi $
orbitals of the carbon atoms. These effects are small for nanotubes with a
sufficiently large diameter.

\subsection{One dimensional rings}

Here, we start with the simplest case $D=1$ having a compact dimension of
the length $L=aN_{c}$ and with the phase in the periodicity condition $%
\alpha _{1}=\alpha $. As we have noted, this case can be considered as a
model of a toroidal nanotube in the limit when the length of one of the
compact dimensions is small compared to the other (for the investigation of
persistent currents in toroidal carbon nanotubes within the framework of the
tight-binding approximation, see Refs. \cite{Lin98}).\footnote{%
Of course, the FC and current density in toroidal nanotubes can also be
considered for general values of the lengths $L_{1}$ and $L_{2}$, by using
the formulas given above for the model $D=2$ and $p=0$.} For simplicity we
consider the case of zero chemical potential when the charge density
vanishes. The zero value for the chemical potential is predicted by simple
tight-binding calculations for the hexagonal lattice of a single graphene
sheet (see, for example, \cite{Sait98}). By using the formulas given in
previous sections, the generalization for the case of a nonzero chemical
potential is straightforward (for mechanisms of generation of a nonzero
chemical potential see Ref. \cite{Shar04}).

By summing the contributions coming from two sublattices with opposite signs
of $\alpha $ and adding an additional factor 2 which takes into account the
contributions from two spins $s=\pm 1$, for the FC we find%
\begin{equation}
\left\langle \bar{\psi}\psi \right\rangle =-\frac{8x_{c}}{\pi L}%
\sum_{l=1}^{\infty }\sin (2\pi l\alpha )\sin (2\pi l\phi /\phi
_{0})K_{0}(lx_{c})+\frac{4x_{c}}{L}\sum_{l=-\infty }^{+\infty }\sum_{j=\pm 1}%
\frac{j/b_{l}^{(j)}}{e^{b_{l}^{(j)}y_{c}}+1},  \label{FCncD1}
\end{equation}%
with the notations%
\begin{equation}
x_{c}=N_{c}\Delta /\gamma _{F},\;y_{c}=\gamma _{F}/(N_{c}T),  \label{xc}
\end{equation}%
and%
\begin{equation}
b_{l}^{(j)}=\sqrt{\kappa _{l}^{(j)2}+x_{c}^{2}},\;\kappa _{l}^{(j)}=2\pi
(l+j\alpha -\phi /\phi _{0}).  \label{blj}
\end{equation}%
The first term in the right-hand side of Eq. (\ref{FCncD1}) gives the FC at
zero temperature. As it is seen, the quantity $\gamma _{F}/N_{c}$ determines
the characteristic energy scale of the model. For toroidal nanotubes one has
$N_{c}\sim 10^{3}$ and this energy scale is of the order of the meV. The
corresponding characteristic temperature is $T_{c}=\gamma _{F}/(N_{c}k_{%
\mathrm{B}})\approx 30\,\mathrm{K}$.

\begin{figure}[tbph]
\begin{center}
\epsfig{figure=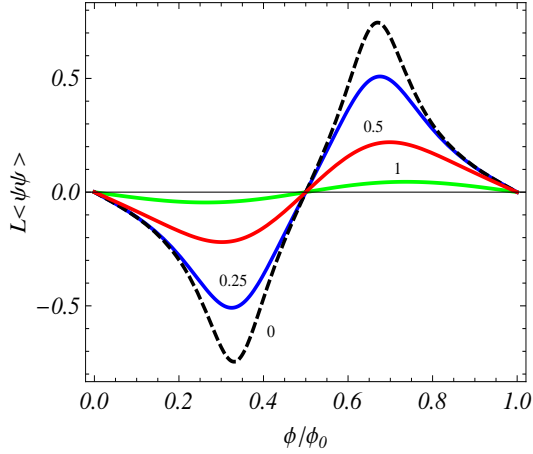,width=7.cm,height=6.cm}
\end{center}
\caption{FC as a function of the magnetic flux for $\protect\alpha =1/3$ and
$x_{c}=0.5$. The numbers near the curves correspond to the value of $TN_{c}/%
\protect\gamma _{F}$.}
\label{fig4}
\end{figure}

An alternative expression is obtained by making use of Eq. (\ref{FC2}):%
\begin{equation}
\left\langle \bar{\psi}\psi \right\rangle =-\frac{16x_{c}}{\pi L}%
\sideset{}{'}{\sum}_{n=0}^{\infty }(-1)^{n}\sum_{l=1}^{\infty }\sin (2\pi
l\alpha )\sin (2\pi l\phi /\phi _{0})K_{0}(x_{c}\sqrt{l^{2}+n^{2}y_{c}^{2}}),
\label{FCncD11}
\end{equation}%
where the zero-temperature part is given by the term $n=0$. From the
asymptotic analysis in Section \ref{sec:FC} we see that at high
temperatures, $T\gg \gamma _{F}/N_{c}$, the FC is exponentially suppressed:%
\begin{equation}
\left\langle \bar{\psi}\psi \right\rangle \approx -\frac{16x_{c}\sin (2\pi
l\alpha )}{\pi L}\sin (2\pi l\phi /\phi _{0})e^{-\pi /y_{c}}.
\label{FCncHighT}
\end{equation}%
An exponential suppression takes place also for large values of the energy
gap: $x_{c}\gg 1$. In figure \ref{fig4} we have plotted the FC, given by Eq.
(\ref{FCncD1}), as a function of the magnetic flux for $\alpha =1/3$ and $%
x_{c}=0.5$. The numbers near the curves correspond to the value of $%
TN_{c}/\gamma _{F}$.

Now we turn to the current density. By summing the contributions coming from
two valleys with opposite signs of $\alpha $, from Eq. (\ref{jnu1}), with
the replacements (\ref{Replace}), we get%
\begin{equation}
\left\langle j^{1}\right\rangle =\frac{8ev_{F}x_{c}}{\pi L}%
\sum_{n=1}^{\infty }\cos (2\pi n\alpha )\sin (2\pi n\phi /\phi
_{0})K_{1}(nx_{c})+\frac{4ev_{F}}{L}\sum_{l=-\infty }^{+\infty }\sum_{j=\pm
1}\frac{\kappa _{l}^{(j)}/b_{l}^{(j)}}{e^{b_{l}^{(j)}y_{c}}+1},
\label{j1ncb}
\end{equation}%
where the first term represents the zero-temperature part and has been
already given in Ref. \cite{Bell10}. Alternatively, from Eq. (\ref{jnu6})
one obtains an equivalent representation
\begin{equation}
\left\langle j^{1}\right\rangle =\frac{16ev_{F}x_{c}^{2}}{\pi L}%
\sideset{}{'}{\sum}_{n=0}^{\infty }(-1)^{n}\sum_{l=1}^{\infty }l\cos (2\pi
l\alpha )\sin (2\pi l\phi /\phi _{0})f_{1}(x_{c}\sqrt{l^{2}+n^{2}y_{c}^{2}}).
\label{j1nc1b}
\end{equation}%
At high temperatures, we have the asymptotic expression
\begin{equation}
\left\langle j^{1}\right\rangle \approx \frac{16ev_{F}}{Ly_{c}}\cos (2\pi
n\alpha )\sin (2\pi n\phi /\phi _{0})e^{-\pi /y_{c}},  \label{j1ncHighT}
\end{equation}%
with the exponential suppression.

\begin{figure}[tbph]
\begin{center}
\begin{tabular}{cc}
\epsfig{figure=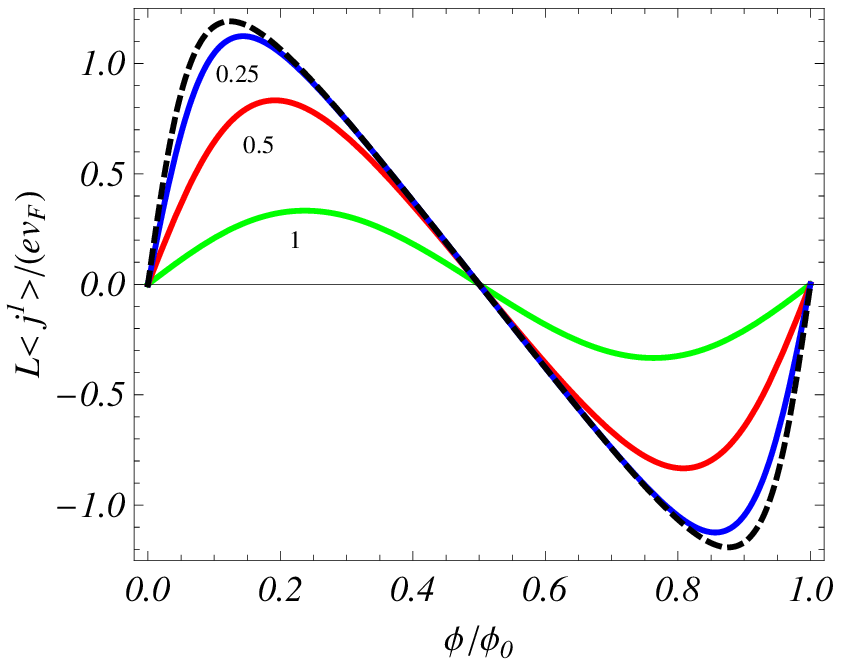,width=7.cm,height=6.cm} & \quad %
\epsfig{figure=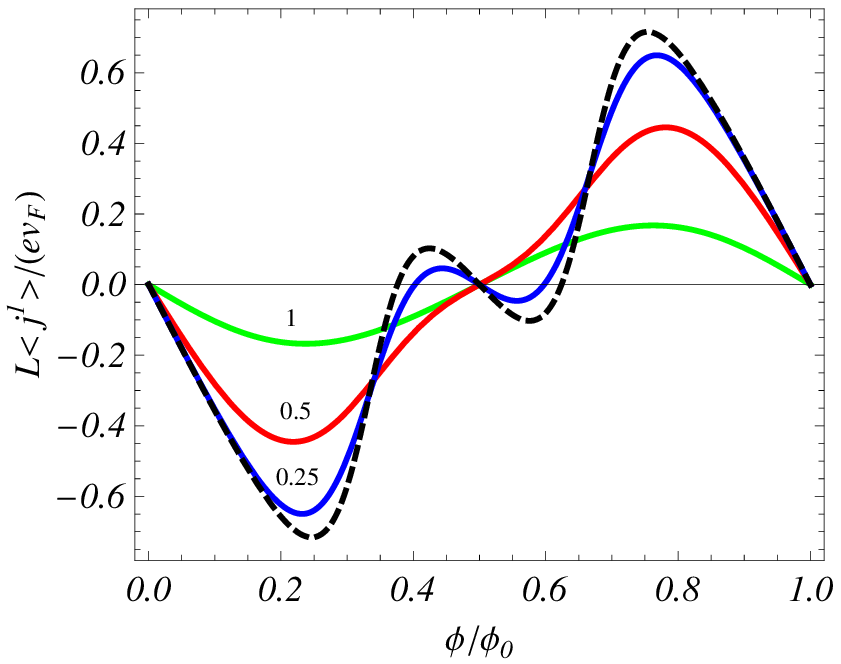,width=7.cm,height=6.cm}%
\end{tabular}%
\end{center}
\caption{Current density as a function of the magnetic flux for $\protect%
\alpha =0$ (left panel) and $\protect\alpha =1/3$ (right panel).
The graphs are plotted for $x_{c}=0.5$ and the curves are labelled
with the value of $TN_{c}/\protect\gamma _{F}$.} \label{fig5}
\end{figure}

For a zero gap energy, the expressions for the current density are reduced to%
\begin{eqnarray}
\left\langle j^{1}\right\rangle &=&\left\langle j^{1}\right\rangle _{T=0}+%
\frac{4ev_{F}}{L}\sum_{l=-\infty }^{+\infty }\sum_{j=\pm 1}\frac{\kappa
_{l}^{(j)}/|\kappa _{l}^{(j)}|}{e^{|\kappa _{l}^{(j)}|y_{c}}+1}  \notag \\
&=&\left\langle j^{1}\right\rangle _{T=0}+\frac{16ev_{F}}{\pi L}%
\sum_{n=1}^{\infty }(-1)^{n}\sum_{l=1}^{\infty }l\sin (2\pi l\phi /\phi _{0})%
\frac{\cos (2\pi l\alpha )}{l^{2}+n^{2}y_{c}^{2}}  \label{j1ncm0}
\end{eqnarray}%
where for the zero-temperature part one has
\begin{equation}
\left\langle j^{1}\right\rangle _{T=0}=\frac{2ev_{F}}{L}\sum_{j=\pm 1}%
\mathcal{I}(\phi /\phi _{0}+j\alpha ).  \label{j1ncT0}
\end{equation}%
In Eq. (\ref{j1ncT0}), we have defined the function%
\begin{equation}
\mathcal{I}(x)=\frac{2}{\pi }\sum_{n=1}^{\infty }\frac{\sin (2\pi nx)}{n}%
=\left\{
\begin{array}{cc}
1-2\{x\}, & x>0 \\
2|\{x\}|-1, & x<0%
\end{array}%
\right. ,  \label{Ix}
\end{equation}%
with $\{x\}$ being the fractional part of $x$. In figure \ref{fig5} we have
plotted the current density, in units of $ev_{F}/L$, as a function of the
magnetic flux for $\alpha =0$ (left panel) and $\alpha =1/3$ (right panel)
with a fixed value of the gap corresponding to $x_{c}=0.5$. The numbers near
the curves are the values of $TN_{c}/\gamma _{F}$. The persistent currents
in normal metal rings with the radius $\approx 10^{3}\,\mathrm{nm}$ have
been recently experimentally observed \cite{Bluh09}. Measurements agree well
with calculations based on a model of non-interacting electrons.

\subsection{FC and the current in cylindrical nanotubes}

\begin{figure}[tbph]
\begin{center}
\epsfig{figure=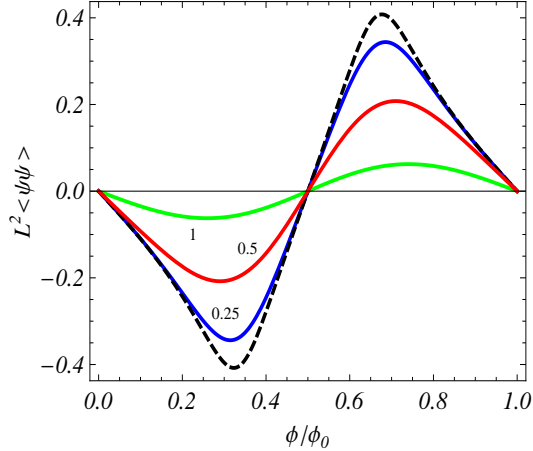,width=7.cm,height=6.cm}
\end{center}
\caption{FC in a carbon nanotube as a function of the magnetic flux for $%
\protect\alpha =1/3$ and $x_{c}=0.5$. The numbers near the curves correspond
to the values of $TN_{c}/\protect\gamma _{F}$.}
\label{fig6}
\end{figure}

The contributions from separate valleys to the FC and the current density
along the compact dimension for a cylindrical nanotube are obtained
specifying in the formulas above $p=q=1$ and $N=2$. Combining these
contributions in accordance with Eq. (\ref{EVs}) and adding a factor 2
corresponding to the spin states $s=\pm 1$, by taking into account Eq. (\ref%
{Replace}), for the FC we get the following representations:%
\begin{equation}
\left\langle \bar{\psi}\psi \right\rangle =-\frac{4x_{c}}{\pi L^{2}}%
\sum_{n=1}^{\infty }\bigg[\frac{\sin (2\pi n\alpha )}{ne^{nx_{c}}}\sin (2\pi
n\phi /\phi _{0})+(-1)^{n}\sum_{l=-\infty }^{+\infty }\sum_{j=\pm
1}jK_{0}(nb_{l}^{(j)}y_{c})\bigg],  \label{FCcn1}
\end{equation}%
with the same notations as in Eq. (\ref{FCncD1}), and
\begin{equation}
\left\langle \bar{\psi}\psi \right\rangle =-\frac{8x_{c}}{\pi L^{2}}%
\sideset{}{'}{\sum}_{n=0}^{\infty }(-1)^{n}\sum_{l=1}^{\infty }\sin (2\pi
l\alpha )\sin (2\pi l\phi /\phi _{0})\frac{e^{-x_{c}\sqrt{%
l^{2}+n^{2}y_{c}^{2}}}}{\sqrt{l^{2}+n^{2}y_{c}^{2}}}.  \label{FCcn2}
\end{equation}%
At high temperatures one has the asymptotic behavior%
\begin{equation}
\left\langle \bar{\psi}\psi \right\rangle \approx -\frac{16x_{c}\sin (2\pi
l\alpha )}{\pi L^{2}\sqrt{2y_{c}}}\sin (2\pi l\phi /\phi _{0})e^{-\pi
/y_{c}}\ .  \label{FCntHightT}
\end{equation}%
For single-walled cylindrical nanotubes $N_{c}\sim 10$ and for the
characteristic energy scale one has $\gamma _{F}/N_{c}\sim 0.2\,\mathrm{eV}$%
. The corresponding characteristic temperature $T_{c}=\gamma _{F}/(N_{c}k_{%
\mathrm{B}})\approx 3\cdot 10^{3}\,\mathrm{K}$. For $T\ll k_{\mathrm{B}%
}T_{c} $, the finite temperature corrections are small and the contribution
of the vacuum expectation value dominates. In figure \ref{fig6} we have
plotted the FC, given by Eq. (\ref{FCcn1}), as a function of the magnetic
flux for $\alpha =1/3$ and $x_{c}=0.5$. The numbers near the curves
correspond to the value of $TN_{c}/\gamma _{F}$.

\begin{figure}[tbph]
\begin{center}
\begin{tabular}{cc}
\epsfig{figure=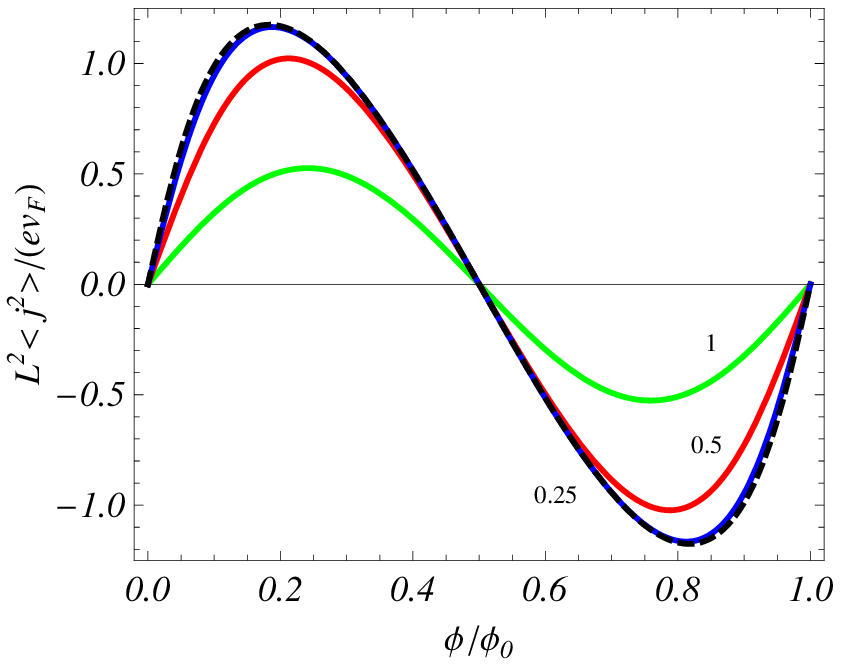,width=7.cm,height=6.cm} & \quad %
\epsfig{figure=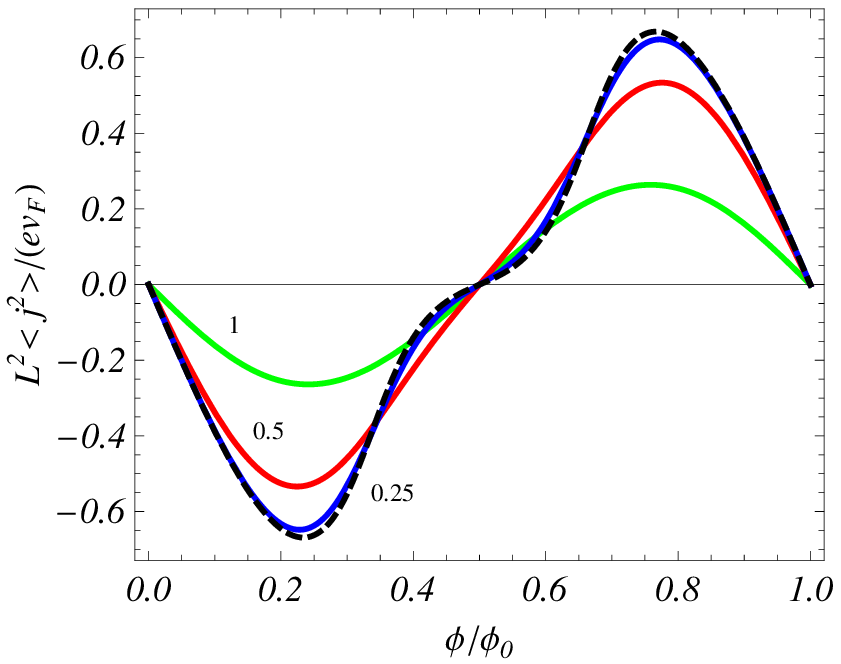,width=7.cm,height=6.cm}%
\end{tabular}%
\end{center}
\caption{Current density along the compact dimension of a cylindrical
nanotube as a function of the magnetic flux for $\protect\alpha =0$ (left
panel) and $\protect\alpha =1/3$ (right panel). The graphs are plotted for $%
x_{c}=0.5$ and the numbers near the curves correspond to the values of $%
TN_{c}/\protect\gamma _{F}$.}
\label{fig7}
\end{figure}

Now let us consider the current density. By summing the contributions coming
from two separate valleys with opposite signs of $\alpha $, again, we find
two representations:%
\begin{equation}
\left\langle j^{2}\right\rangle =\frac{4ev_{F}}{\pi L^{2}}\sum_{n=1}^{\infty
}\bigg[ \cos (2\pi n\alpha )\sin (2\pi n\phi /\phi _{0})\frac{1+nx_{c}}{%
n^{2}e^{nx_{c}}}-(-1)^{n}\sum_{l=-\infty }^{+\infty }\sum_{j=\pm 1}\kappa
_{l}^{(j)}K_{0}(nb_{l}^{(j)}y_{c})\bigg] ,  \label{j2nc}
\end{equation}%
and%
\begin{equation}
\left\langle j^{2}\right\rangle =\frac{8ev_{F}}{\pi L^{2}}\sideset{}{'}{\sum}%
_{n=0}^{\infty }(-1)^{n}\sum_{l=1}^{\infty }l\sin (2\pi l\phi /\phi _{0})%
\frac{\cos (2\pi l\alpha )(1+x_{c}\sqrt{l^{2}+n^{2}y_{c}^{2}})}{%
(l^{2}+n^{2}y_{c}^{2})^{3/2}e^{x_{c}\sqrt{l^{2}+n^{2}y_{c}^{2}}}}.
\label{j2nc2}
\end{equation}%
In the absence of a gap energy one has $x_{c}=0$, and in Eq. (\ref{j2nc}) we
take $b=|\kappa _{l}^{(j)}|$. At high temperatures the current density is
exponentially suppressed:%
\begin{equation}
\left\langle j^{2}\right\rangle \approx \frac{16e\cos (2\pi n\alpha )}{\sqrt{%
2}L^{2}y_{c}^{3/2}}\sin (2\pi n\phi /\phi _{0})e^{-\pi /y_{c}} \ .
\label{j2ncHighT}
\end{equation}%
In figure \ref{fig7} we displayed the current density, in units of $ev_{F}/L$%
, as a function of the magnetic flux for $\alpha =0$ (left panel) and $%
\alpha =1/3$ (right panel) with fixed value of the gap corresponding to $%
x_{c}=0.5$. The numbers near the curves correspond to the values of $%
TN_{c}/\gamma _{F}$.

\section{Conclusion}

\label{sec:Conc}

In this paper we have considered the combined effects of finite temperature
and nontrivial topology on the FC and on the expectation values of charge
and current densities for a massive fermionic field. Along compact
dimensions, quasiperiodicity conditions (\ref{PerCond}) are imposed with
arbitrary phases $\alpha _{l}$. Twisted and untwisted periodicity
conditions, most often discussed in the literature, are special cases. In
addition, we have assumed the presence of a constant gauge field which gives
rise to Aharonov-Bohm-like effects on the expectation values. The effects of
nontrivial phases in the boundary conditions and of the gauge field appear
in the form of the combination $\tilde{\alpha}_{l}$, given by Eq. (\ref{epsn}%
).

Under the condition $\left\vert \mu \right\vert <\varepsilon _{0}$, with $%
\varepsilon _{0}$ being the lowest energy in the energy spectrum of a
fermionic particle (see Eq. (\ref{epsmin})), the FC is given by Eq. (\ref%
{FCInt1}), where $\left\langle \bar{\psi}\psi \right\rangle _{0}$ is the FC
at zero temperatures and the second term on the right presents the finite
temperature contribution. The FC is an even periodic function of $\tilde{%
\alpha}_{l}$ with the period equal to unity. In particular, it is a periodic
function of the fluxes enclosed by compact dimensions with the period equal
to the flux quantum. As a limiting case, from the general result we have
derived the expression for the FC in the topologically trivial Minkowski
spacetime, Eq. (\ref{FCM}). If the length of one of the compact dimensions,
say the $l$th dimension, is much smaller than the other length scales, the
behavior of the FC crucially depends on the value of the parameter $\tilde{%
\alpha}_{l}$. Assuming that $|\tilde{\alpha}_{l}|<1/2$, for $\tilde{\alpha}%
_{l}\neq 0$ and for small values of $L_{l}$, the FC is suppressed by the
factor $e^{-2\pi |\tilde{\alpha}_{l}|\beta /L_{l}}$. In the case $\tilde{%
\alpha}_{l}=0$, to the leading order the FC coincides with $N\left\langle
\bar{\psi}\psi \right\rangle _{(p,q-1)}/(N_{D-1}L_{l})$, where $\left\langle
\bar{\psi}\psi \right\rangle _{(p,q-1)}$ is the FC in the $(D-1)$%
-dimensional space of topology $R^{p}\times (S^{1})^{q-1}$. At low
temperatures, the thermal corrections are suppressed by the factor $%
e^{-(\varepsilon _{0}-|\mu |)/T}$ (see Eq. (\ref{FClow})).

An alternative expression for the FC, Eq. (\ref{FC3}), is obtained by using
the analytic continuation of the generalized zeta function (\ref{Zeta})
provided by the Chowla-Selberg formula. A representation of the FC, Eq. (\ref%
{FC3b}), convenient in the investigation of the high temperature limit is
obtained from Eq. (\ref{FC3}) with the help of formula (\ref{SumHighT}). At
high temperatures, the dominant contribution comes from the terms with $n=0$
and $n=-1$ and the effects induced by the nontrivial topology are suppressed
by the factor $e^{-\pi T L_{\mathrm{min}}}$, with $L_{\mathrm{min}}$ being
the smallest length of the compact dimensions. In the high-temperature
expansion of the FC, the leading terms coincides with that for the FC in the
topologically trivial Minkowski spacetime with the asymptotic behavior given
by Eq. (\ref{FCMhighT}). The representation of the FC, Eq. (\ref{FC4}),
containing more detailed explicit information, is obtained by applying a
variant of the Abel-Plana summation formula (\ref{AP}). The second term in
the right-hand side of Eq. (\ref{FC4}) presents the part in the FC induced
by the compactification of the $l$-th dimension. This information is not
explicit in the previous two representations.

In the case $|\mu |>$ $\varepsilon _{0}$, assuming that $\mu >0$, the
expression for the finite temperature part in the FC coming from the
antiparticles remains the same, whereas in the part coming from the
particles the spectral ranges $\varepsilon _{\mathbf{n}_{q}}\leqslant
\varepsilon _{\mathbf{n}_{q}^{(0)}}$ and $\varepsilon _{\mathbf{n}%
_{q}}>\varepsilon _{\mathbf{n}_{q}^{(0)}}$ should be treated separately and
the expression for the FC takes the form (\ref{FCmu}). Now, at zero
temperature one has Eq. (\ref{FCmut0}), where the second contribution in the
right-hand side comes from the particles which occupy the states with $%
\varepsilon _{\mathbf{n}_{q}}<\mu $. At high temperatures, the leading term
in the asymptotic expression for the FC remains the same, and it coincides
with that in the topologically trivial Minkowski spacetime.

The expectation value of the charge density is investigated in Section \ref%
{sec:Charge}. In the case $\left\vert \mu \right\vert <\varepsilon _{0}$,
the zero temperature charge density vanishes, and the effects induced by the
finite temperature are given by Eq. (\ref{C2}). The charge density is an
even periodic function of the phases $\tilde{\alpha}_{l}$ with the period
equal to unity. Besides, it is an odd function of the chemical potential $%
\mu $. For large values of the lengths of compact dimensions, in the leading
order we recover the result for the topologically trivial Minkowski bulk,
Eq. (\ref{j0M}). At low temperatures the induced charge is exponentially
small (see Eq. (\ref{Clow})). An equivalent representation for the charge
density, Eq. (\ref{CZ2}), is obtained by making use of the related zeta
function analytic continuation. At high temperatures the part in the charge
density coming from the nontrivial topology is exponentially suppressed, and
the leading term in the asymptotic expansion is given by the Minkowskian
part, Eq. (\ref{j0MhighT}). Besides, another representation for the charge
density, Eq. (\ref{CAP}), is obtained by using the Abel-Plana formula. This
representation separates the part due to the compactification of the $l$th
compact dimension.

For $\left\vert \mu \right\vert >\varepsilon _{0}$ and assuming that $\mu >0$%
, the contribution to the charge density coming from antiparticles remain
the same, and the total charge density is given by Eq. (\ref{C0mu}) (or
equivalently by Eq. (\ref{Q})). In this case the charge density at zero
temperature is presented as Eq. (\ref{j0T0}). The latter is related to the
number of states with the energies smaller than $|\mu |$ by a simple formula
$\left\langle j^{0}\right\rangle _{T=0}=e\mathcal{N}_{\leqslant \mu }$. For
a fixed value of the charge, the formulas for the charge density determine
the chemical potential as a function of the charge, the temperature and the
volume of the compact subspace. At high temperatures this function decays
with the leading term (\ref{muHightT}). Decreasing the temperature, the
chemical potential increases and its value at the zero temperature is
determined by Eq. (\ref{j0T0}).

An interesting effect induced by the nontrivial topology is the appearance
of the nonzero current density along compact dimensions. The current density
along the $\nu $-th compact dimension, given by Eq. (\ref{jnu1}) for $%
\left\vert \mu \right\vert <\varepsilon _{0}$, is an even periodic function
of the phases $\tilde{\alpha}_{l}$, $l\neq \nu $, with the period equal to
unity, and an odd periodic function of the phase $\tilde{\alpha}_{\nu }$. In
particular, the current density vanishes for twisted and untwisted fields in
the absence of the gauge field. The current density is an even function of
the chemical potential. At low temperatures it coincides with the
corresponding result al zero temperature, given by Eq. (\ref{jnT0}), for $%
\left\vert \mu \right\vert <\varepsilon _{0}$, up to exponentially small
terms. At high temperatures, the thermal corrections to the current density
along the $\nu $th compact dimension are suppressed by the factor $e^{-\pi
TL_{\nu }}$. This behavior is in sharp contrast with the high-temperature
asymptotic of the current density in the case of a scalar field. For the
latter, the current density linearly grows with the temperature. For $\mu
>\varepsilon _{0}$, the contribution of the particles to the current density
with the energies $\varepsilon _{\mathbf{n}_{q}}<\mu $ should be considered
separately and the corresponding expression takes the form (\ref{jnimu}).
Now the current density at zero temperature receives the contributions from
both virtual and real particles. The latter is given by the second term in
the right-hand side of Eq. (\ref{jnumuT0}). The asymptotic behavior at high
temperatures remains the same. We have also investigated the current density
along compact dimensions for a fixed value of the charge. A numerical
example for a $D=3$ model with a single compact dimension is presented in
figure \ref{fig3}.

For odd spacetime dimensions, with an irreducible representation of the
Clifford algebra, the mass term breaks $C$-invariance in $D=4n$, $P$%
-invariance in $D=4n,4n+2$, and $T$-invariance in $D=4n+2$. In the absence
of magnetic fields, these symmetries are restored in the model with two
fermionic fields realizing two inequivalent representations of the Clifford
algebra. These two fields can be combined in a single $2N$-component spinor
with the Lagrangian density (\ref{Loddb}). The respective FC and the current
densities are obtained by combining the corresponding results for the upper
and lower components of this spinor in the form of Eq. (\ref{EVs}). As an
application of the general results we have considered the $D=2$ model used
for the effective field theoretical description of low-energy degrees of
freedom of the electrons in a graphene sheet. For carbon nanotubes the
corresponding topology is $R^{1}\times S^{1}$. Depending on the chiral
vector of the nanotube, for the phases in the quasiperiodicity conditions
one has $\alpha _{l}=0,\pm 1/3$, and the phases have opposite signs for the
upper and lower components of the 4-spinor. Compactifying the direction
along the nanotube axis, one gets toroidal carbon nanotubes with the
topology of a two-torus for the effective Dirac theory. As a simple model of
a toroidal nanotube, we have considered a one-dimensional ring. For
simplicity, in both cases of cylindrical and toroidal topologies we have
assumed the zero chemical potential. In this case the charge density
vanishes. On the base of the formulas for general $D$, the generalization of
the corresponding expressions of the FC and current density in nanotubes for
the case of a nonzero chemical potential is straightforward. \

\section*{Acknowledgments}

ERBM thanks Conselho Nacional de Desenvolvimento Cient\'{\i}fico e Tecnol%
\'{o}gico (CNPq) for partial financial support. AAS was supported by State
Committee Science MES RA, within the frame of the research project No. SCS
13-1C040. This work was partially supported by the ERC Advanced Grant no.
226455 \textit{``Supersymmetry, Quantum Gravity and Gauge Fields''}~(\textit{%
SUPER\-FIELDS}).

\end{document}